\documentclass[fleqn,usenatbib]{mnras}

\usepackage{newtxtext,newtxmath}
\usepackage{xcolor}

\usepackage[T1]{fontenc}
\usepackage{ae,aecompl}

\usepackage{graphicx}	
\usepackage{amsmath}	
\usepackage{amssymb}	

\usepackage{amsmath,amssymb, graphics}
\usepackage{pgf,tikz}
\usepackage{bm}
\usepackage{graphicx}
\usepackage{epstopdf}
\usepackage{float}

\newcommand{\Mp}{M_{\rm p}}
\newcommand{\Mj}{{\rm M}_{\rm Jup}}
\newcommand{\Rp}{R_{\rm p}}
\newcommand{\Rj}{{\rm R}_{\rm Jup}}
\newcommand{\ap}{a_{\rm p}}
\newcommand{\ep}{e_{\rm p}}
\newcommand{\J}{J_2}
\newcommand{\Ms}{M_\star}
\newcommand{\Msun}{{\rm M}_\odot}
\newcommand{\bTst}{{\bm T}_\star}

\newcommand{\bTsp}{{\bm T}_{\rm p}}

\newcommand{\bl}{\bm {\hat l}}
\newcommand{\blp}{\bm {\hat l}_{\rm p}}
\newcommand{\bs}{\bm {\hat s}}
\newcommand{\rin}{r_{\rm in}}
\newcommand{\rout}{r_{\rm out}}
\newcommand{\rL}{r_{\rm L}}

\newcommand{\Om}{\Omega}
\newcommand{\om}{\omega}
\newcommand{\bgp}{\beta_{\rm p}}
\newcommand{\bb}{\beta}

\newcommand{\bg}{\beta}

\newcommand{\Dg}{\Delta}

\newcommand{\bcdot}{{\bm \cdot}}

\newcommand{\be}{\begin{equation}}
\newcommand{\ee}{\end{equation}}

\title[Stability of Circumplanetary Disks]{The Structure and Stability of Extended, Inclined Circumplanetary Disk or Ring Systems} 

\author[Speedie \& Zanazzi]{
Jessica Speedie,$^{1,2}$\thanks{jspeedie@cita.utoronto.ca (JS)}
J.J. Zanazzi$^{1}$\thanks{jzanazzi@cita.utoronto.ca (JJZ)}
\\
$^{1}$Canadian Institute for Theoretical Astrophysics, University of Toronto, 60 St. George Street, Toronto, Ontario, M5S 1A7, Canada\\
$^{2}$School of Interdisciplinary Science, McMaster University
Hamilton ON L8S 4M1, Canada
}

\date{Accepted 2020 July 7. Received 2020 June 16; in original form 2019 November 26}

\pubyear{2020}

\begin{document}
\label{firstpage}
\pagerange{\pageref{firstpage}--\pageref{lastpage}}
\maketitle

\begin{abstract}
Large dips in the brightness for a number of stars have been observed, for which the tentative explanation is occultation of the star by a transiting circumplanetary disk or ring system.  In order for the circumplanetary disk/rings to block the host star's light, the disk must be tilted out of the planet's orbital plane, which poses stability problems due to the radial extent of the disk required to explain the brightness dip durations.  This work uses $N$-body integrations to study the structure and stability of circumplanetary disk/ring systems tilted out of the planet's orbital plane by the spinning planet's mass quadrupole.  Simulating the disk as a collection of test particles with orbits initialized near the Laplace surface (equilibrium between tidal force from host star and force from planet's mass quadrupole), we find that many extended, inclined circumplanetary disks remain stable over the duration of the integrations ($\sim 3-16 \, {\rm Myr}$).  Two dynamical resonances/instabilities excite the particle eccentricities and inclinations: the Lidov-Kozai effect which occurs in the disk's outer regions, and ivection resonance which occurs in the disk's inner regions.  Our work places constraints on the maximum radial extent of inclined circumplanetary disk/ring systems, and shows that gaps present in circumplanetary disks don't necessarily imply the presence of exomoons.
\end{abstract}



\section{Introduction}

The detection and characterization of circumplanetary disks and/or rings around giant planets may soon be at the frontier of exoplanetary science.  If a giant planet is embedded in its natal protoplanetary disk, a young gaseous circumplanetary disk is most detectable from its near-infared emission \citep{Szulagyi(2019)nearIR}, but may also be detectable from its continuum \citep{Szulagyi(2018)cont} and polarized scattered light \citep{Szulagyi(2019)scatt} emission.  Recently, a circumplanetary disk around the Jovian planet PDS 70b has been detected, inferred by the disk's excess infared emission, and H$\alpha$ and Br$\gamma$ line emission from accretion onto the planet \citep{Christiaens(2019)}.

For more evolved systems, one can detect disks/rings around transiting planets using photometry, when the rings transit before (after) the planet during ingress (egress) \cite[e.g.][]{Schneider(1999),barnes2004transit,Ohta(2009),zuluaga2015novel,SandfordKipping(2019),ReinOfir(2019)}.  Numerous searches for planetary rings have been conducted, and while most have been inconclusive \cite[e.g.][]{Brown(2001),Heising(2015),Santos(2015)}, one possible detection of an exoring system has been made around the planet KIC 10403228 \citep{Aizawa(2017)}.  These searches have targeted planets that may host rings orbiting out to distances a few times larger than the host planet's radius, similar to the rings around Saturn.

In contrast to these searches for relatively compact exoring systems, a number of tentative detections of more extended exoplanetary disks have been made around dipper stars.  Dipper stars typically lie in young stellar associations (ages $\lesssim 100 \, {\rm Myr}$), and display periodic, quasi-periodic, or aperiodic brightness variations (along with many other stars in the cluster) which cannot be explained by intrinsic stellar activity, but are thought to be due to material in the surrounding environment blocking some of the star's flux.  These stellar brightness dips have been attributed to many astrophysical phenomena, including occultations from dusty material near the inner edge of circumstellar disks \citep[e.g.][]{Cody(2014),McGinnis(2015),Ansdell(2016),Hedges(2018)}, transiting exocomets \citep[e.g.][]{Rappaport(2018), Ansdell(2019)}, and tidally disrupted disks from binary interactions \citep[e.g.][]{Rodriguez(2018)}.  Of interest for this work are the ``deep dipper'' stars 1 SWASP J140747-354542 \citep{mamajek2012planetary}, PDS 110 \citep{Osborn(2017),Osborn(2019)}, VVV-WIT-07 \citep{Saito(2019)}, and EPIC 204376071 \citep{Rappaport(2019)}, with large ($\gtrsim 30-80\%$) brightness dips.  One proposed explanation for the deep dipper brightness fluctuations is a transit by a planet hosting an extended, optically thick circumplanetary\footnote{Throughout this work, we will refer to these disks as circumplanetary, but note the secondarys' masses may be sub-stellar ($\gtrsim 13 \, \Mj$)} disk/ring\footnote{We will refer to the structure surrounding the planet as a disk and/or ring system interchangably throughout this work} system.  In order to explain the depth and duration of these large brightness decreases, the circumplanetary disk/ring system must extend out to distances comparable to the planet's Hill radius (e.g. \citealt{mamajek2012planetary}, see also Sec.\ref{sec:j1407b}).


The most studied of these deep dippers is the star 1 SWASP J140747-354542 (hereafter J1407).  In 2012, \citet{mamajek2012planetary} presented photometry from the SuperWASP (Super Wide Angle Search for Planets) database \citep{pecaut2013intrinsic} of J1407, a young ($\sim$16 Myr old) K5 pre-main sequence star at $\sim$130 pc in the Sco-Cen OB Association. The star's light curve exhibited a complex series of deep eclipses that lasted $\sim54$ days around the month of April 2007. The authors interpreted the light curve of J1407 as the transit of a circumplanetary disk surrounding an unseen substellar companion, J1407b. They fit the gross features of the nightly mean SuperWASP photometry with a simple ring model and proposed that we had detected one of the first circumplanetary disks outside our Solar system \citep{mamajek2012planetary}. 

The J1407b system has since been the subject of a number of subsequent studies, which have refined the simple ring model to accurately fit the reduced light curve \citep{van2014analysis, kenworthy2015modeling}, placed constraints on the mass and period of the companion \citep{kenworthy2014mass, mentel2018constraining}, constrained the radial extent of the disk \citep{rieder2016constraints}, and explored whether the rapid flux variations in J1407's light curve can be attributed gaps opened by mean motion resonances with nearby moons \citep{sutton2019mean}. The best fitting ring model presented in \citet{mamajek2012planetary} is flat, inclined with respect to the companion's orbital plane by 24.2$^{\circ}$, and very extended, with an outer radius of $\rout \sim 0.6 \, {\rm au}$.

While the J1407b circumplanetary disk hypothesis (along with PDS 110, VVV-WIT-07, and EPIC 204376071) remains unconfirmed (see Sec. \ref{sec:j1407b} for a discussion), the question of the dynamical stability of extended circumplanetary disks is pertinent to the upcoming discovery of more exoring candidates. It is likely that our first detections will be of extended ring systems due to their larger transit signal. For exorings to produce a transit signal at all, the disk must be inclined with respect to the planet's orbital plane, and the rings must be stable to gravitational perturbations from the host star over the system's lifetime \citep[e.g.][]{SchlichtingChang(2011),zanazzi2016extended}.  Without any stabilizing forces, the gravitational influence of the host star will destroy the rings over a timescale much shorter than the system's age \citep{sucerquia2017anomalous}.

The rings and many satellites of the four giant planets in our Solar system are inclined with respect to the planets' orbital planes (and are aligned with the planets' equators) due to the fact that the planets are spinning and oblique. The competition between the interior quadrupole potential from a spinning planet's equatorial bulge and the external tidal potential from the host star determines an equilibrium surface surrounding the planet: the ``Laplace surface'' \citep{laplace1805vol}.  In this work, we explore the dynamical stability of the Laplace surface around an oblate and oblique planet with $N$-body integrations, assuming the particles which compose the disk are massless.  Our goal is to probe the stability of collisionless circumplanetary debris disks, applying our results to the tentative exoring systems already detected, with an eye towards constraining the properties of exoring systems which may be detected in the future.  In Section \ref{sec:simulations}, we describe our circumplanetary disk model in detail and outline how we perform our $N$-body integrations. We present our main results in Section \ref{sec:results}, and discuss their implications in Section \ref{sec:discussion}.  Section~\ref{sec:conclusions} summarizes our results and draws our main conclusions.


\section{Simulations}
\label{sec:simulations}

\subsection{System Setup}
\label{subsec:background}

Consider a planet with mass $\Mp$ and radius $\Rp$ in an orbit with semi-major axis $\ap$, eccentricity $\ep$, and orbital angular momentum unit vector $\blp$, around a host star of mass $\Ms$.  The planet's rotation makes it oblate, giving it a mass quadrupole coefficient $J_2$, and has a spin orbital angular momentum unit vector $\bs$.  The planet's obliquity $\bgp$ is the angle between $\blp$ and $\bs$ ($\cos \bgp = \blp \bcdot \bs$).  

\begin{figure}
	\includegraphics[width=\columnwidth]{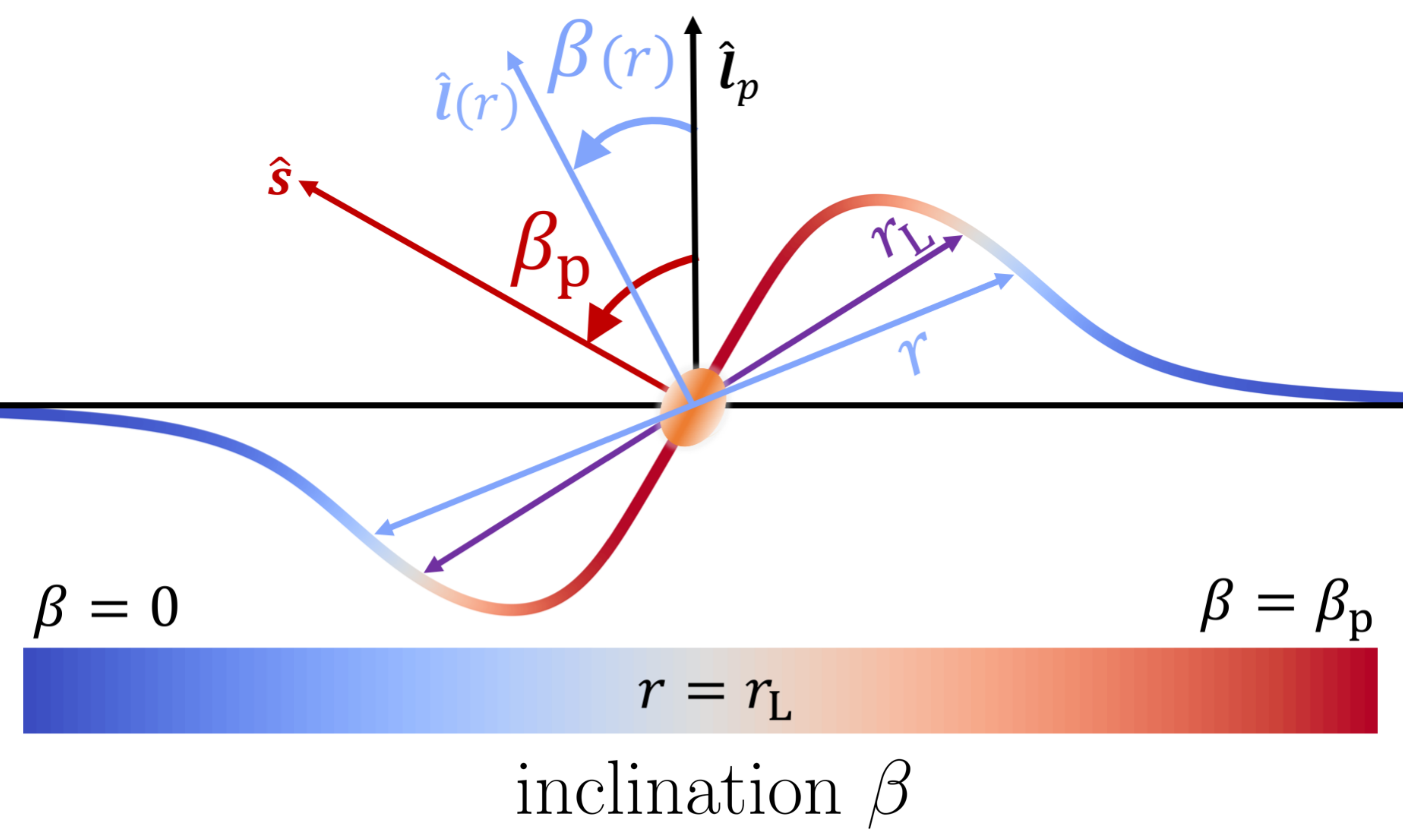}
    \caption{A schematic for our circumplanetary disk setup.  The spinning oblate planet (orange) with spin orbital angular momentum unit vector $\bs$ is inclined to the planet's orbital angular momentum unit vector $\blp$ by an angle $\bgp$ (planet's obliquity).  Individual satellite particles (assumed massless) orbit a distance $r$ from the planet with orbital angular momentum unit vectors $\bl = \bl(r)$ on the equilibrium/Laplace surface, inclined to $\blp$ by a angle $\bg$ (test particle's obliquity; eq.~\ref{eq:beta(r)}).  The Laplace radius $\rL$ (eq.~\ref{eq:rL}) denotes where the Laplace surface warp is (approximately) highest.  We adopt the color scheme displayed by the color bar for $\bg$ values throughout this work, unless otherwise indicated.
    }
    \label{fig:schematic}
\end{figure}

\begin{figure}
	\includegraphics[width=\columnwidth]{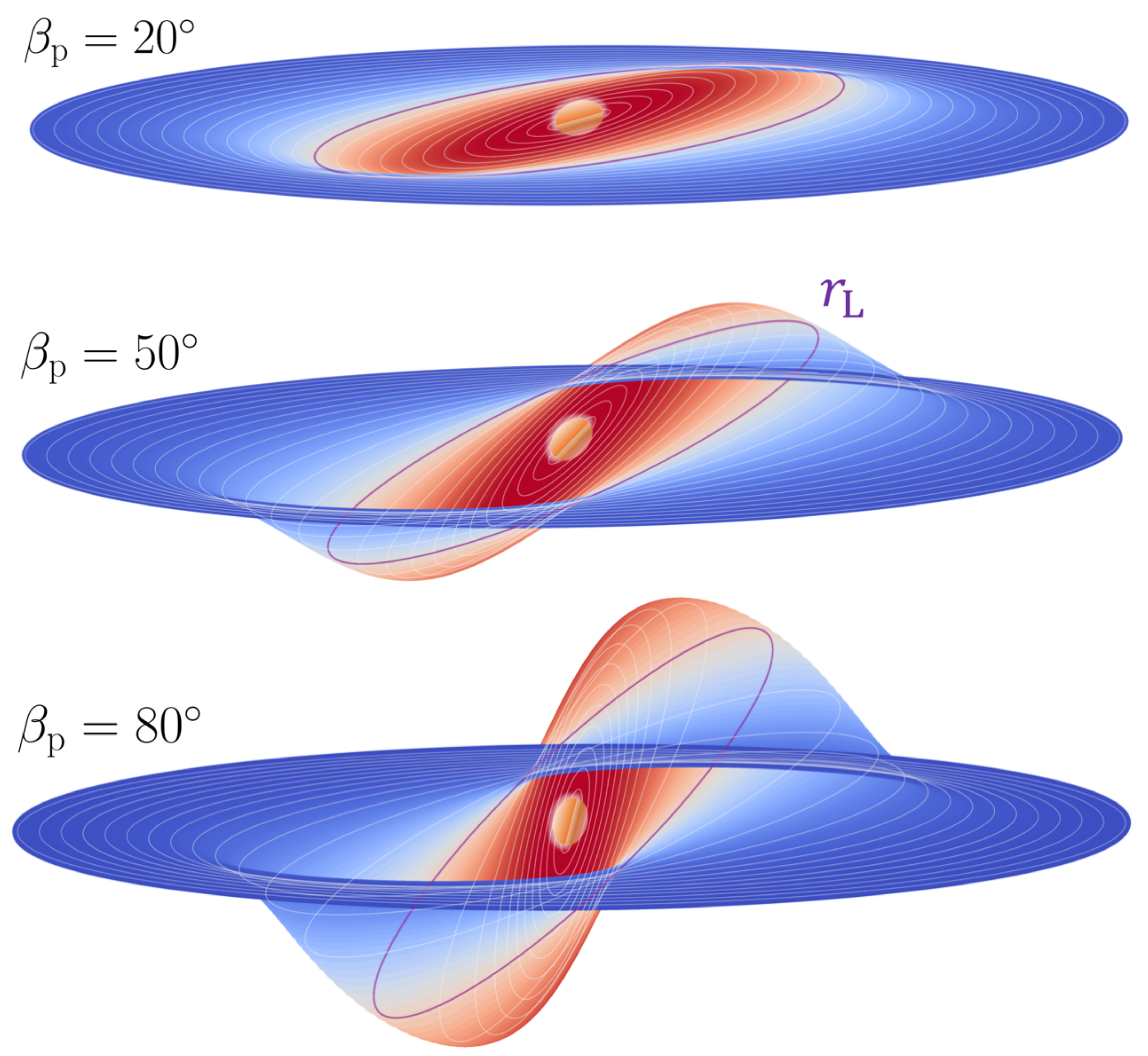}
    \caption{Three-dimensional visualizations of equilibrium/Laplace surfaces for the planetary obliquities $\bgp$ indicated.  The Laplace radius $\rL$ (eq.~\ref{eq:rL}) is depicted by the purple circle. 
    }
    \label{fig:surface}
\end{figure}

Orbiting the planet is a massless ring system with inner (outer) truncation radius $\rin$ ($\rout$), with $\bl = \bl(r,t)$ specifying the orbital angular momentum unit vector of a ring with semi-major axis $r$ at time $t$.  Two torques act on the rings.  The first is the tidal torque from the distant host star:
\begin{equation}
    \label{eq:bTst}
    \bTst = \frac{3 G \Ms r^2}{4\ap^3} (\bl \cdot \blp) (\bl \times \blp),
\end{equation}
where we have averaged $\bTst$ over the planet's orbit, and assumed $r \ll \ap$ and the ring eccentricity $e \ll 1$.  The second is the torque from the spinning oblate planet:
\begin{equation}
    \label{eq:bTsp}
    \bTsp = \frac{3 G \Mp \Rp^2 \J}{2 r^3} (\bl \cdot \bs) (\bl \times \bs).
\end{equation}
The equilibrium surface (Laplace surface) is set by torque balance ($\bTst + \bTsp = 0$).  Since the only stable equilibrium for $\bl$ driven solely by $\bTst$ ($\bTsp$) is to be aligned with $\blp$ ($\bs$), the equilibrium value for $\bl(r)$ under the combined influence of $\bTst$ and $\bTsp$ must lie in the plane spanned by $\blp$ and $\bs$.  Defining $\bg(r)$ as the inclination between $\blp$ and $\bs$ ($\cos \bg[r] = \blp \bcdot \bl[r]$), one can solve for the equilibrium profile \citep[e.g.][]{tremaine2009satellite, Tamayo(2013)}:
\begin{equation}
    \label{eq:beta(r)}
    \bb (r) = \bgp - \frac{1}{2} \arctan{\bigg[ \frac{\sin{2\bgp }}{ \cos{2\bgp} + (\rL / r)^5 } \bigg]},
\end{equation}
where
\begin{align}
    \label{eq:rL}
    \centering
    \rL = &  \left[ 2\, \Big(1 - \ep^2\Big)^{3/2} \, \J \, \Rp^2 \,  \ap^3 \,  \frac{\Mp}{\Ms} \,    \right]^{1/5} \nonumber \\
    =\ & 0.02225 \, \Big(1 - \ep^2\Big)^{3/10} \, \Big(\frac{\J}{0.1}\Big)^{1/5} \, \Big(\frac{\Rp}{1 \, \Rj}\Big)^{2/5} \,  \Big(\frac{\ap}{5 \text{ au}}\Big)^{3/5} \nonumber \\
    & \times  \Big(\frac{\Mp}{1 \, \Mj}\Big)^{1/5} \,  \Big(\frac{\Msun}{1\, \Ms}\Big)^{1/5}  \text{ au}
\end{align}
is the Laplace radius, which denotes roughly where the torques from the distant host star and oblate planet balance (ignoring geometrical factors), and where the Laplace surface warp is strongest.

Figure~\ref{fig:schematic} displays a schematic of the Laplace surface.  Interior to the Laplace radius ($r < \rL$), the torque acting on rings from the oblate planet dominates, and ring annuli are roughly aligned with the planet's equatorial plane ($\bg \approx \bgp$).  Exterior to the Laplace radius ($r > \rL$), the host star's tidal torque dominates, and ring annuli are roughly aligned with the planet's orbital plane ($\bg \approx 0$).  Three-dimensional visualizations of the Laplace surface are displayed in Figure~\ref{fig:surface}, for multiple obliquity values $\bgp$.  We see the radial extent of the disk's misaligned outer region is roughly given by $\rL$, hence any transit signal from the circumplanetary ring system will have an effective outer radius $\rL$.

\subsection{Test Particle Integrations}
\label{subsec:rebound}


We perform our integrations using the \texttt{WHFAST} integrator \citep{rein2015whfast} in the open-source \texttt{REBOUND} \textit{N}-body package \citep{rein2012rebound}.  We distribute 1000 test particles linearly in semi-major axis $r$ from an inner truncation radius $\rin = 0.2 \, \rL$ to an outer truncation radius $\rout = 2 \, \rL$.  We divide the 1000 particles by semi-major axis into 50 segments of 20 particles each, with segment widths $\Delta r = 0.0018 \, \rL$.  We integrate each segment separately, taking the segment's time-step to be $7\%$ of the inner-most particle's orbital period within the segment.  With this method, we are able to integrate the outer $\sim800$ particles for the full 16 Myr (the age of the J1407 system), but integrate the inner $\sim200$ particles to $\sim$3-15 Myr due to computational constraints. We note that 16 Myr corresponds to $\sim$13-82 $\times 10^6$ circumplanetary orbits for the outer $\sim$800 particles, and 3-15 Myr corresponds to $\sim$68-86 $\times 10^6$ orbits for the inner $\sim$200 particles.  A break-down of the total integration time by particle and segment for one of our integrations is provided in Table~\ref{tab:appendix} of Appendix \ref{sec:app:integration}.

The host star is taken to have a mass of $\Ms=1 \, \Msun$, and the planet's mass $\Mp=10 \, \Mj$, radius $\Rp=1 \, \Rj$, and semi-major axis $\ap=5 \, {\rm au}$, with a mass quadrupole coefficient $J_2=0.1$.  We include the force acting on the test particles from the oblate planet by adding a custom-defined additional force, using the publically available example ``\texttt{J2\_force}'' \texttt{C} code in the \texttt{REBOUND} repositories.  The $J_2$ value chosen in our integrations is likely high, since the most oblate planet in our solar system (Saturn) has $\J = 0.0162$ \citep{murray1999solar}.  We take $J_2 = 0.1$ to maximize the radial extent of the inclined region of the circumplanetary disk: since $r_L$ has a weak dependence on $J_2$ ($\rL \propto J_2^{1/5}$), our results do not depend sensitively on the $J_2$ value.  We perform a set of six simulations, varying the planet's obliquity to be $\bgp=20^{\circ}$, $50^{\circ}$, or $80^{\circ}$, and its eccentricity to be $\ep = 0$ or $0.5$.  These values are motivated by the tentative J1407b system, but are broadly applicable to extended and inclined circumplanetary disk systems in general (see Sec.~\ref{sec:discussion} for discussion).   

\begin{figure}
	\includegraphics[width=\columnwidth]{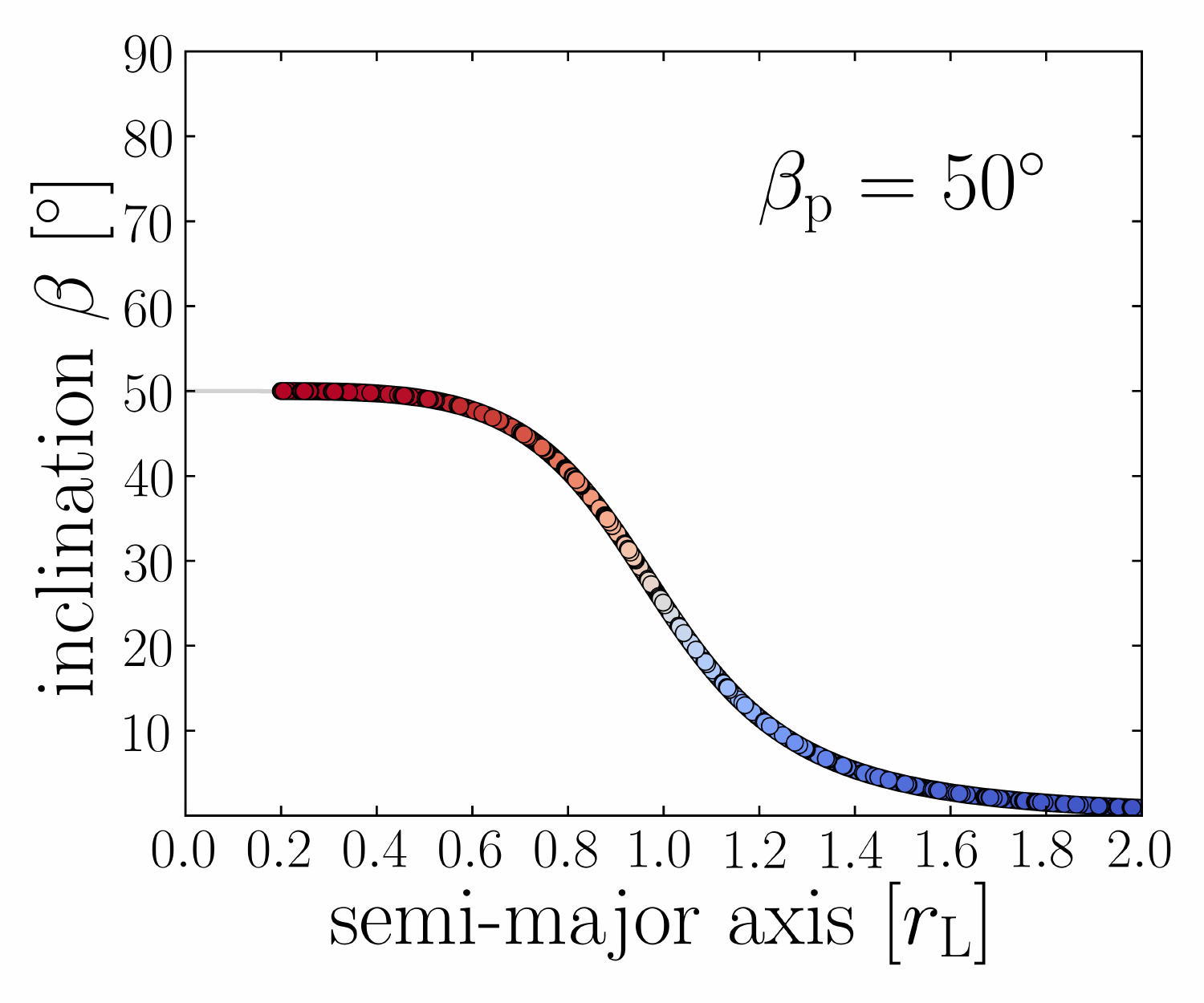}
    \caption{The inclinations $\bg$ of 1000 test particles over the course of the integration ($\sim 3-16 \, {\rm Myr}$, see text for details), initialized exactly on the Laplace surface (eq.~\ref{eq:beta(r)}).  Empty black circles denote test particle initial conditions, while colored dots track the test particle's evolution.  We sample the integration once every 100 years, and randomly pick 1 in 5 of these points to display.  The planet's obliquity $\bgp = 50^\circ$ and planet's orbital eccentricity $\ep = 0$.}
    \label{fig:surfacesim}
\end{figure}

A test of our simulation setup is displayed in Figure~\ref{fig:surfacesim}.  Particles are initialized with no eccentricity, and with orbital planes which lie exactly on the Laplace surface (inclinations $\bg$ given by eq.~\ref{eq:beta(r)}).  The particle inclinations remain constant over the entire course of the simulation ($\sim 3-16 \, {\rm Myr}$), with negligible changes in the particles' longitude of ascending nodes $\Om$ and semi-major axis $r$.  The stability of particles on the Laplace surface confirms we have implemented the planet's $J_2$ and host star's tidal forces correctly, and that equation~\eqref{eq:beta(r)} describes the disk's equilibrium warp profile.

To probe the stability of a circumplanetary disk near the Laplace surface, we draw the test particle orbital parameters from distributions which differ slightly from their exact equilibrium values (Fig.~\ref{fig:surfacesim}).  We draw the test particle eccentricity $e$ from a Rayleigh distribution with mode 0.01, and argument of pericenter $\om$ from a uniform distribution from $-\pi$ to $\pi$ rad.  Test particle inclinations are drawn from normal distributions centered on the equilibrium warp profile given by equation~\eqref{eq:beta(r)}, with a standard deviation of 0.01 rad.  Our results are detailed in the next section.

\section{Results}
\label{sec:results}

\begin{figure*}
	\includegraphics[width=18cm]{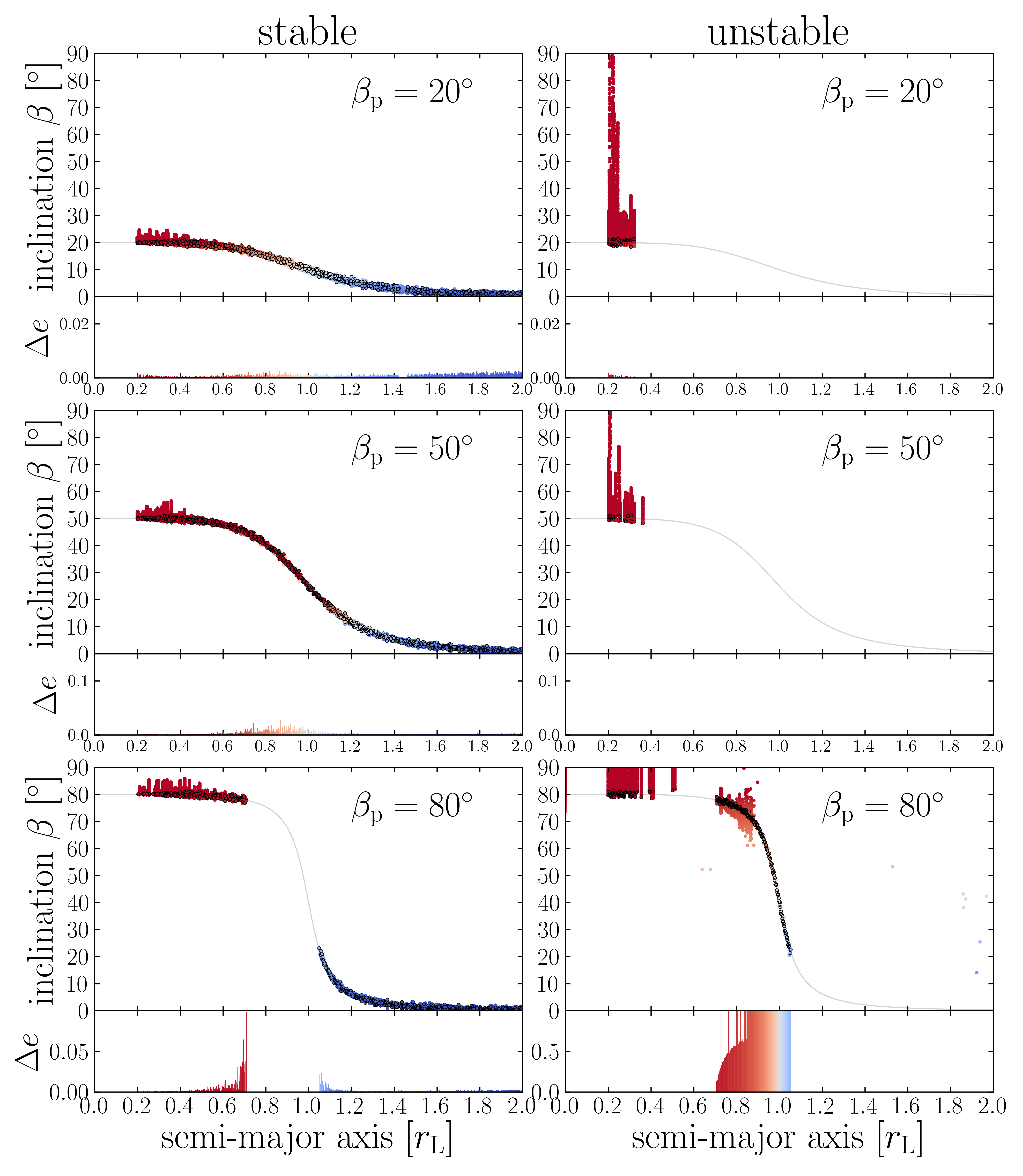}
    \caption{N-body integrations of test particles initialized near the Laplace surface (see Sec.~\ref{subsec:rebound} for details), for a planet on a circular orbit ($\ep=0$), for the planetary obliquities $\bgp$ as indicated. The top plot in each panel shows inclinations $\bb$ of $N=1000$ particles over the full course of each integration, with empty  black circles indicating their initial positions around the Laplace surface (grey line) and coloured points tracking their evolution. The bottom plot in each panel shows the change in particle eccentricity $\Dg e = \max[e(t)]$ over the course of the integration.  Particles with orbits that remain near the Laplace surface ($|\Delta \bb| = |\bb[t]-\bb[0]| < 0.1$ rad and $e[t] < 0.1$) are categorized as stable (left panel), and particles that become significantly inclined to the Laplace surface ($|\Delta \bb| > 0.1$ rad) or develop significant eccentricities ($e > 0.1$) as unstable (right panel).  Each integration is sampled once every 100 years ($\sim$11 planet orbits), and we randomly select 1 in 5 points in each sample to plot. Here, $\rL = 0.035 \, {\rm au} = 0.048 \, r_{\rm H}$, where $r_{\rm H} = \ap (1-\ep)[\Mp/(3 \Ms)]^{1/3}$ is the Hill radius. }
    \label{fig:betapanel-ep0}
\end{figure*} 



\begin{figure*}
	\includegraphics[width=18cm]{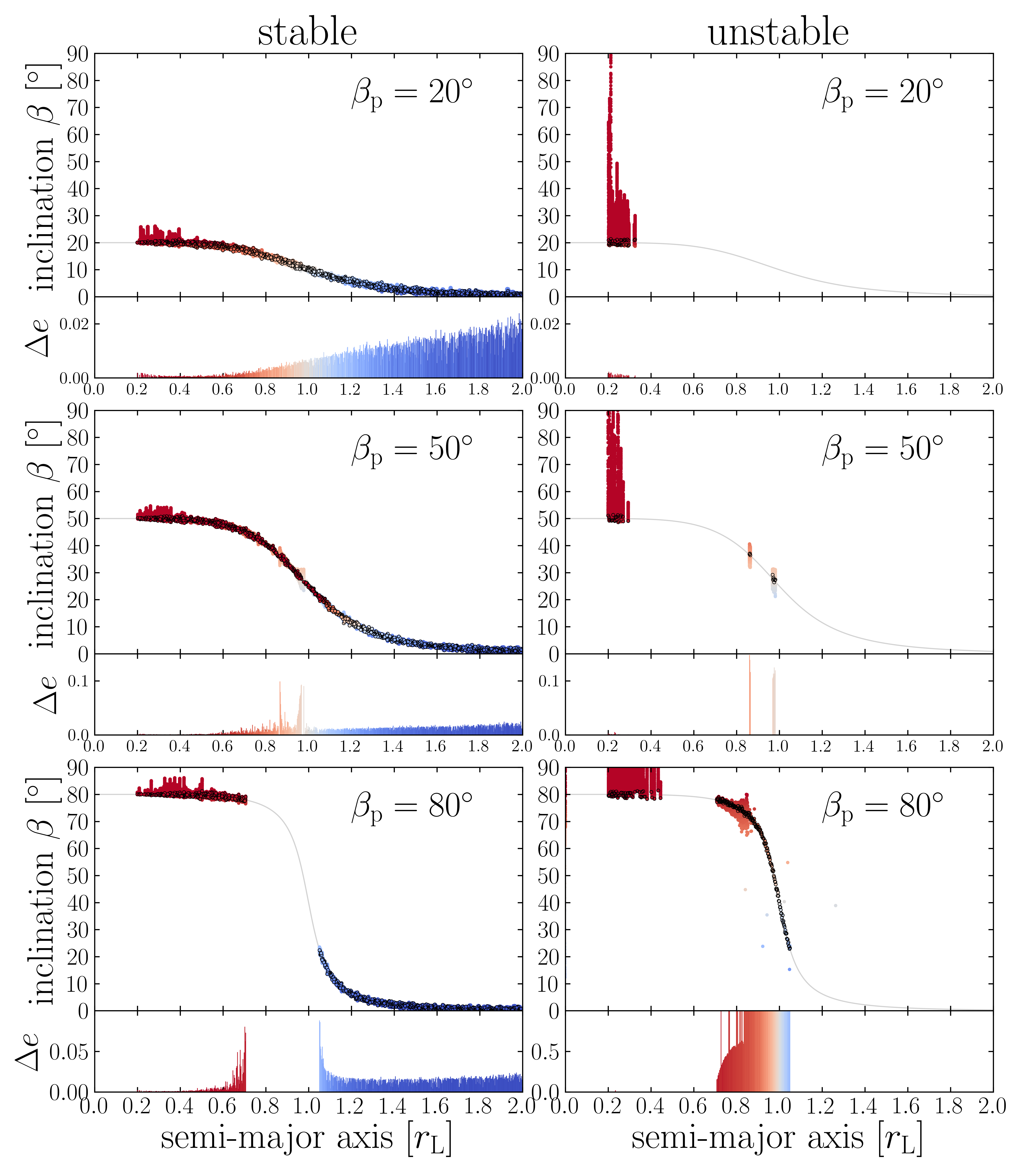}
    \caption{Same as Figure~\ref{fig:betapanel-ep0}, except the planet lies on an eccentric orbit ($\ep=0.5$). Here, $\rL = 0.032 \, {\rm au} = 0.088 \, r_{\rm H}$, where $r_{\rm H} = \ap (1-\ep)[\Mp/(3 \Ms)]^{1/3}$ is the Hill radius.
    }
    \label{fig:betapanel-ep05}
\end{figure*}

In this section, we present the results of our N-body integrations for a collection of test particles initialized near the Laplace surface, integrating to $\sim 3-16 \, {\rm Myr}$ depending on the particle's orbital period (see Sec.~\ref{subsec:rebound} for discussion). 

After running the integrations for a set of system parameters, we group particles into one of two categories: stable or unstable. If the test particle's inclination stays close to the Laplace surface ($|\Dg \bg[t]| = |\bg[t] - \bg[0]| < 0.1 \, {\rm rad}$), and if its orbit remains close to circular ($e[t]<0.1$), we classify the particle as stable. If the test particle becomes significantly inclined with respect to the Laplace surface ($|\Dg \bg[t]| > 0.1 \, {\rm rad}$) or develops a significant eccentricity ($e[t] > 0.1$), we classify the particle as unstable. Misalignment with the Laplace plane can also occur if a particle's longitude of ascending node $\Omega$ moves out of the Laplace plane. Figure \ref{fig:omega-v-time} shows the vast majority of stable particles have $\Om$ values close to the Laplace surface's longudude of ascending node ($\Om = -\pi/2$).

Figures~\ref{fig:betapanel-ep0} and~\ref{fig:betapanel-ep05} plot the stable and unstable test particles' inclination $\bg$ and semi-major axis $r$ trajectories, with the change in particle eccentricity $\Delta e$, for a planet on a circular ($\ep = 0$; Fig.~\ref{fig:betapanel-ep0}) and eccentric ($\ep = 0.5$; Fig.~\ref{fig:betapanel-ep05}) orbit, varying the planet's obliquity $\bgp$.  For low to moderate planet obliquities ($\bgp = 20^\circ, 50^\circ$), $N \gtrsim 950$ out of $1000$ particles remain stable over the entire course of the integration, for circular and eccentric planetary orbits.  Most particles have orbits which remain close to the Laplace surface over the disk's entire annular extent ($0.2 \, \rL \lesssim r \lesssim 2.0 \, \rL$).  For high planet obliquities ($\bg = 80^\circ$), only test particles close to the planet ($r \lesssim 0.71 \, \rL$) or past the Laplace radius ($r \gtrsim 1.05 \, \rL$) remain stable over the course of the integration.

We expect every instability/resonance observed in our integrations to be secular over the test particle's orbit.  Comparing the particle's mean-motion $n = \sqrt{G \Mp/r^3}$ to the planet's $n_{\rm p} = \sqrt{G \Ms/\ap^3}$, we see
\begin{align}
    \frac{n}{n_{\rm p}} = \frac{89}{(1-\ep^2)^{9/20}} &\left( \frac{0.1}{J_2} \right)^{1/5} \left( \frac{\Mp}{\Mj} \right)^{1/5} \left( \frac{1 \, \Msun}{\Ms} \right)^{1/5}
    \nonumber \\
    &\times \left( \frac{\ap}{5 \, {\rm au}} \right)^{3/5} \left( \frac{1 \, {\rm R}_{\rm Jup}}{\Rp} \right)^{3/5} \left( \frac{r_{\rm L}}{r} \right)^{3/2}.
\end{align}
Since $n \gg n_{\rm p}$ when $r \lesssim {\rm few} \ \rL$, we expect mean-motion resonances to be unimportant.  As expected, all stable particles in our integrations have negligible changes in their initial $r$ values, with semi-major axis variations $|r(t)-r(0)| \lesssim 0.01 \, r(0)$.  Unstable particles may have significant changes in their $r$ values when the particle becomes unbound to or crashes into the planet, or from numerical effects when $e$ approaches unity.

Two main instabilities are observed in our integrations.  The most extended instability in $r$ is a Lidov-Kozai (LK) instability which occurs for the $\bgp = 80^\circ$ integration when particles orbit near the Laplace radius ($0.71 \, \rL \lesssim r \lesssim 1.05 \, \rL$), originally predicted to occur by \cite{tremaine2009satellite} when $\bgp > 68.875^\circ$.  The second instability occurs in the disk's inner regions ($r \lesssim 0.3-0.5 \, \rL$), regardless of the planet's $\bgp$ value.  The instabilities occur regardless of the planetary orbit's eccentricity value (Figs.~\ref{fig:betapanel-ep0} \&~\ref{fig:betapanel-ep05}).  To confirm these instabilities are robust to the integration time step, we re-performed our integrations with a time-step set at 1\% the innermost particle's period in each disk segment for $10^4$ years, and observed the instabilities were qualitatively the same.  We discuss both instabilities in greater detail in Sections~\ref{subsec:LK} and~\ref{subsec:ivec}.

For the planet on an eccentric orbit ($\ep = 0.5$, Fig.~\ref{fig:betapanel-ep05}) with an obliquity $\bgp = 50^\circ$, we note an additional instability not present in the similar integration for a planet on a circular orbit (Fig.~\ref{fig:betapanel-ep0}), which occurs on two narrow regions around $r \approx 0.86 \, \rL$ and $r \approx 0.98 \, \rL$.  These instabilities excite test particle inclinations and eccentricities ($\Dg \bg > 0.1 \, {\rm rad}$ and $e > 0.1$), but to relatively modest values ($\Dg \bg \lesssim 0.2 \, {\rm rad}$ and $e \lesssim 0.2$).  This instability may be due to octopole-level perturbations from the distant host star \citep[e.g.][]{Naoz(2011),Katz(2011),Li(2014)}, but we do not discuss this instability any further in this work.

In our integrations with a planet on an eccentric orbit, the projection of the planet's spin axis $\bs$ onto the planet's orbital plane lies perpendicular to the planet's pericenter direction.  Our integrations with $\bs$ pointing in the same or opposite direction as the planet's pericenter, which we do not show, find similar results.

\subsection{Lidov-Kozai Instability} 
\label{subsec:LK}

\begin{figure} 
	\includegraphics[width=\columnwidth]{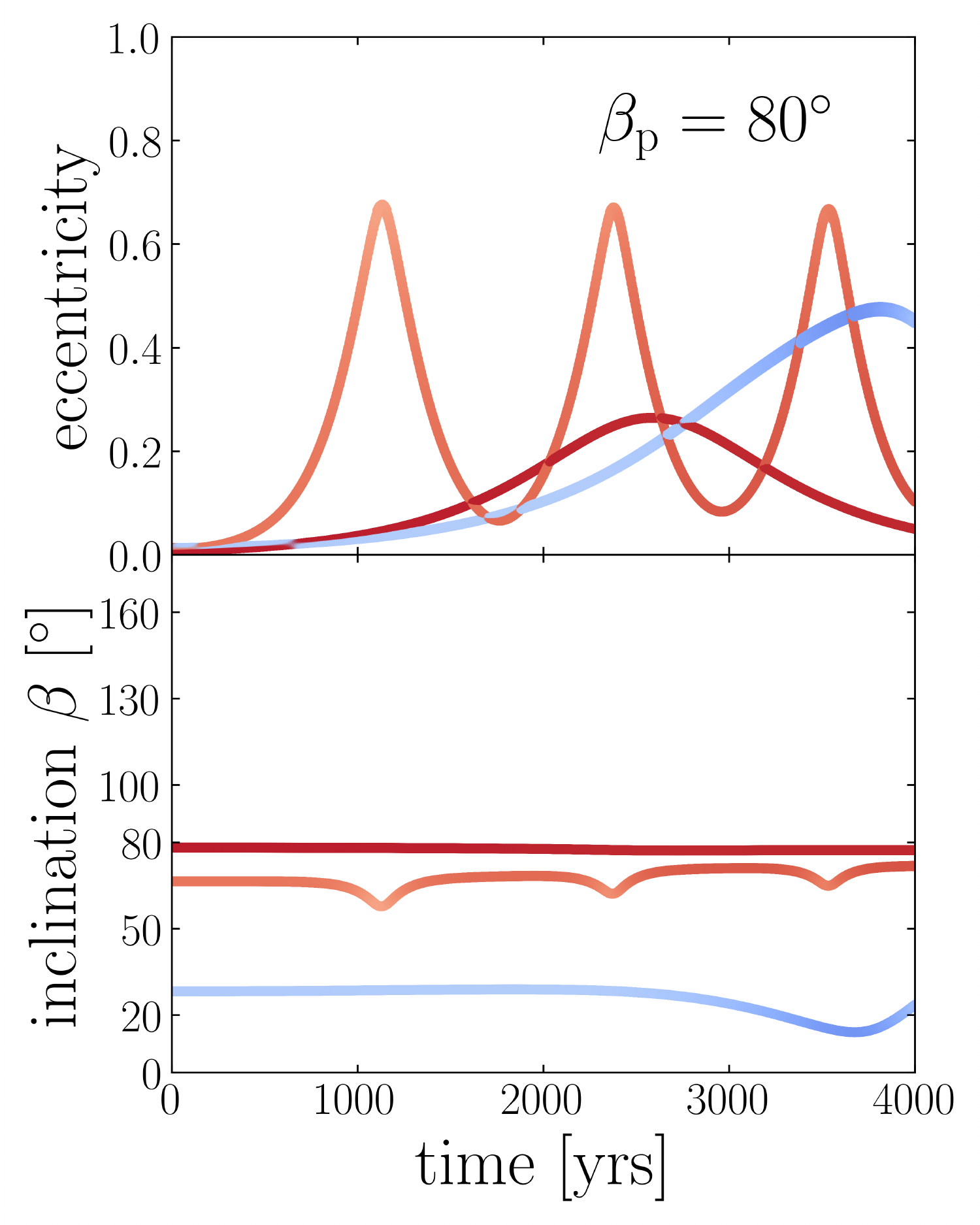}
    \caption{
    Evolution of test particle orbital eccentricity $e$ (top panel) and inclination $\bg$ (bottom panel) for a few representative test particles undergoing the Lidov-Kozai instability, with semi-major axis $a = 0.720 \, \rL$ (dark red), $a = 0.907 \, \rL$ (light red), and $a = 1.028 \, \rL$ (blue, see Fig.~\ref{fig:schematic} for color scheme).  We select particles from the $\ep = 0$, $\bgp = 80^\circ$ integration (Fig.~\ref{fig:betapanel-ep0}, bottom right panel).  We display the test particle evolution for the first 4000 years, sampled twice per year.
    }
    \label{fig:bp>70}
\end{figure}

This subsection analyzes the LK instability seen in our high planetary obliquity ($\bgp = 80^\circ$) integrations in greater detail.  Figure~\ref{fig:bp>70} plots the inclination $\bg$ and eccentricity $e$ evolution of three representative test particles which become LK unstable during the $\bgp = 80^\circ$ integration for a planet on a circular orbit (bottom-right panel of Fig.~\ref{fig:betapanel-ep0}).  We see over secular timescales (${\rm few} \ 10^3 \, {\rm yrs}$), the test particle $\bg$ value decreases when the $e$ value becomes sufficiently large, characteristic of LK oscillations \citep{Lidov(1962),Kozai(1962)}.  Many particles are eventually ejected from the system or crash into the planet, hence the scattering and pile-up at $r \approx 0$ in the bottom right panels of Figures~\ref{fig:betapanel-ep0} and~\ref{fig:betapanel-ep05}. Our plots don't display $\bg$ variations for particles with $r \gtrsim 0.9 \, \rL$,  because the particles become unbound ($e>1$) before the change in $\bg$ can be sampled.

Previous work done by \cite{tremaine2009satellite} predicted test particles near the Laplace surface would be susceptible to the LK instability when $\bgp > 68.875^\circ$, since the host star's tidal torque may overcome the stabilizing influence of the oblate planet's torque, and cause the particle's eccentricity to grow exponentially \citep[see also][]{tremaine2014earth,Martin(2014),Zanazzi(2017)Kozai,LubowOgilvie(2017)}.  \cite{tremaine2009satellite} found the particles which could become LK unstable had $r$ values from about $0.95 \, \rL$ to $1.1 \, \rL$ when $\bgp = 80^\circ$ and $\ep = 0$ (roughly estimated from their Fig.~2).  We find the annular extent of the disk susceptible to the LK instability to be more extended ($0.71 \, \rL \lesssim r \lesssim 1.05 \, \rL$).  Further work is needed to understand why our integrations find LK unstable particles when $r \lesssim 0.95 \, \rL$.

\subsection{Ivection Instability}
\label{subsec:ivec}

\begin{figure}
	\includegraphics[width=\columnwidth]{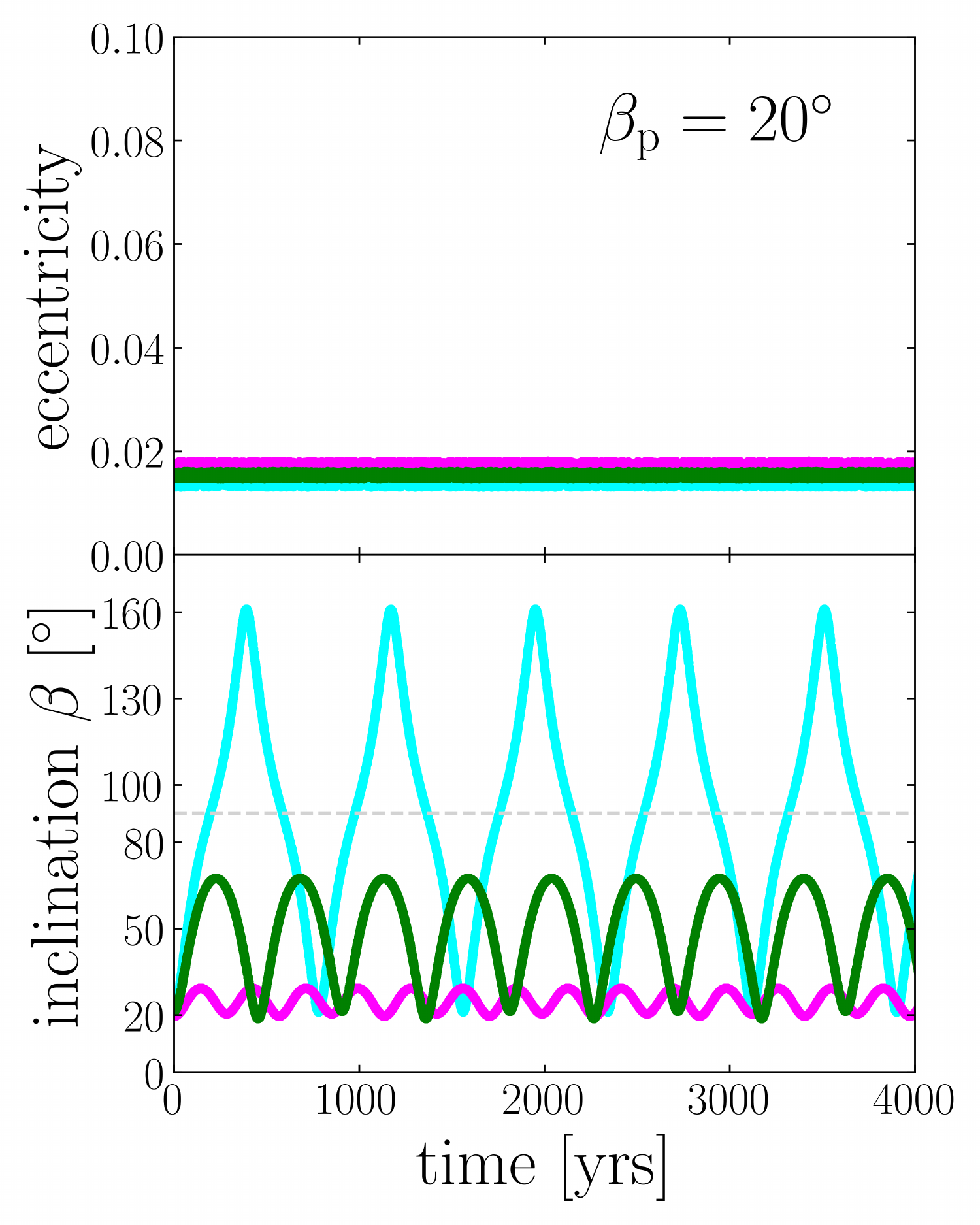}
    \caption{
    Evolution of test particle eccentricities $e$ (top panel) and inclinations $\bg$ (bottom panel) undergoing the ivection instability, from the $\bgp = 20^\circ$, $\ep = 0$ integration (top right panel of Fig. \ref{fig:betapanel-ep0}).  Different colored lines denote the orbital evolution of three seperate particles, with initial orbital elements ($a,e,\bg,\om$) = ($0.200 \, \rL, 0.015, 21.0^\circ,70.8^\circ$) (cyan), ($a,e,\bg,\om$) = ($0.209 \, \rL, 0.017, 19.5^\circ,-29.1^\circ$) (magenta), and ($a,e,\bg,\om$) = ($0.232 \, \rL,0.016,21.3^\circ,-40.7^\circ$) (green), with $\Om$ always initialized to $\Om = -90^\circ$.  The dashed grey line at $90^\circ$ shows the thresh-hold for prograde ($\bg<90^\circ$) vs. retrograde ($\bg > 90^\circ$) orbits.  We display integrations run for 4000 years, sampled twice per year.
    }
    \label{fig:ivec}
\end{figure}

In all of our integrations, test particles close to the host star ($r \lesssim 0.3-0.5 \, \rL$) occasionally have large excitations in their inclinations $\bg$.  This subsection argues the $\bg$ excitation when $r\lesssim 0.3-0.5 \, \rL$ is due to an ivection resonance \citep{xu2019exciting}, and examines the orbital evolution for particles which undergo ivection resonance.

The ivection resonance is very similar to the more well-studied evection resonance \citep[e.g.][]{ToumaWisdom(1998),CukBurns(2004),yokoyama2008evection,frouard2010dynamics,Spalding(2016),xu2016disruption}.  In the context of satellites orbiting giant planets, the evection resonance occurs when the apsidal precession rate $\dot \omega$, driven by either the planet's mass quadrupole or host star's tidal torque, matches the mean-motion of the planet $n_{\rm p} = \sqrt{G \Ms/\ap^3}$.  The effect of this secular-orbital resonance (averaged over the satellites orbit, but not the planet's) is to cause a significant excitation in the satellite's eccentricity $e$, with little corresponding excitation in the satellite's inclination $\bg$.  Ivection resonance is an analogous secular-orbital resonance, but occurs when the precession rate of the satellite's longitude of ascending node $\dot \Om$ matches $n_{\rm p}$, and mainly excites the satellite's $\bg$ rather than $e$.  Although the system setup for the recently discovered ivection resonance \citep{xu2019exciting} is different from this work, consisting of two nearly coplanar planets with a binary companion perturber, there is no reason to expect ivection resonance does not occur for satellites orbiting giant planets.

Within a factor of a few, evection and ivection should occur at similar $r$ values, since the apsidal $\dot \om$ and nodal $\dot \Om$ precession rates driven by the planet's mass quadrupole are similar.  Ignoring geometrical factors, the apsidal/nodal precession rate on the test particle from the oblate planet is \cite[i.e.][]{tremaine2009satellite,tremaine2014earth}
\begin{equation}
    \dot \om|_{\rm s} \sim \dot \Om|_{\rm s} \sim \frac{\J \Rp^2 (G \Mp)^{1/2}}{r^{7/2}}.
\end{equation}
Evection resonance occurs when $\dot \om|_{\rm s} \sim n_{\rm p}$, while ivection resonance occurs when $\dot \Om|_{\rm s} \sim n_{\rm p}$.  Solving for the resonant semi-major axis $r_{\rm res}$ where $\dot \om|_{\rm s} \sim \dot \Om|_{\rm s} \sim n_{\rm p}$, we find
\begin{align}
    r_{\rm res} \sim\  &\J^{2/7} \Rp^{4/7} \ap^{3/7} (\Mp/\Ms)^{1/7}
    \nonumber \\
    =\ &\frac{0.22}{(1-\ep^2)^{3/10}} \left( \frac{\J}{0.1} \right)^{3/35} \left( \frac{\Rp}{1 \, \Rj} \right)^{6/35} \left( \frac{5 \, {\rm au}}{\ap} \right)^{6/35}
    \nonumber \\
    &\times \left( \frac{\Ms}{1 \, \Msun} \right)^{2/35} \left( \frac{1 \, \Mj}{\Mp} \right)^{2/35} \rL.
    \label{eq:r_res}
\end{align}
From estimate~\eqref{eq:r_res}, we see evection and ivection may be important near the disk's inner truncation radius $\rin = 0.2 \, \rL$.

Figure~\ref{fig:ivec} plots the evolution of test particle $e$ and $\bg$ values which undergo ivection resonance.  We see $\bg$ undergoes oscillations, while $e$ remains relatively constant, as expected from ivection \citep{xu2019exciting}.  Although oscillations in $\bg$ are a few tens of degrees for most test particles, some particles display much larger oscillations, ocassionally even becoming retrograde ($\bg > 90^\circ$\, see Figure \ref{fig:ivection-zoom}). Additional visualizations of the variation in particle $\bg$-excitation amongst the ivection-unstable particles are provided Figure \ref{fig:beta-vs-perturbation}. We also ran integrations after decreasing the disk's $r_{\rm in}$ (not shown), and found the disk experienced evection resonance ($e$ excited, $\bg$ relatively constant) at $r \approx 0.1 \, \rL$.  Our exploration of ivection and evection is preliminary, and a more detailed investigation is outside the scope of this work.


\section{Discussion}
\label{sec:discussion}

In this section, we discuss the implications of our work for the J1407b system, as well as future work and theoretical uncertainties.


\subsection{Implications for the J1407b system}
\label{sec:j1407b}

\subsubsection{Are exomoons present?}
\label{sec:exomoons}

The light-curve of J1407 is highly variable and structured, argued to be caused by gaps in J1407b's circumplanetary disk, visible during the system's transit.  Exomoons orbiting J1407b have been invoked to carve these gaps, since this structure should otherwise be erased by internal disk interactions on dynamical timescales \citep{van2014analysis, kenworthy2015modeling}.  In our Solar system, stable and long-lived gaps are created by resonant satellite-disk interactions, and similar processes may be happening in the J1407 system.  \cite{kenworthy2015modeling} even estimated the maximum mass of a moon that could have cleared the largest gap in their ring model (0.0267 au wide at 0.4 au), which they found to be 8 Earth masses. However, \cite{sutton2019mean} tried to reproduce this same gap in $N$-body simulations by placing moons orbiting exterior to the circumplanetary disk, at the locations of mean motion resonances with the gap, and concluded this scenario carved a gap which was too narrow to explain the observations. 

If the disk is not flat, but warped by the competing influence of the tidal torque from the host star and the torque from the spinning planet, we find that secular instabilities can potentially carve gaps in collisionless disks. Ivection resonance could produce highly structured gaps in the inner regions of the disk, and at sufficiently high planetary obliquities ($\bgp > 68.875^\circ$), the secular instability of \cite{tremaine2009satellite} could result in such a large gap near the system's Laplace radius $\rL$ (eq.~\ref{eq:rL}) that there appears to be distinct inner and outer circumplanetary disks.  Neither instability which may carve gaps in the disk requires the presence of exomoons.


\subsubsection{Is the disk warped?}
\label{sec:Obs_rout}

\begin{figure*}
	\includegraphics[width=18cm]{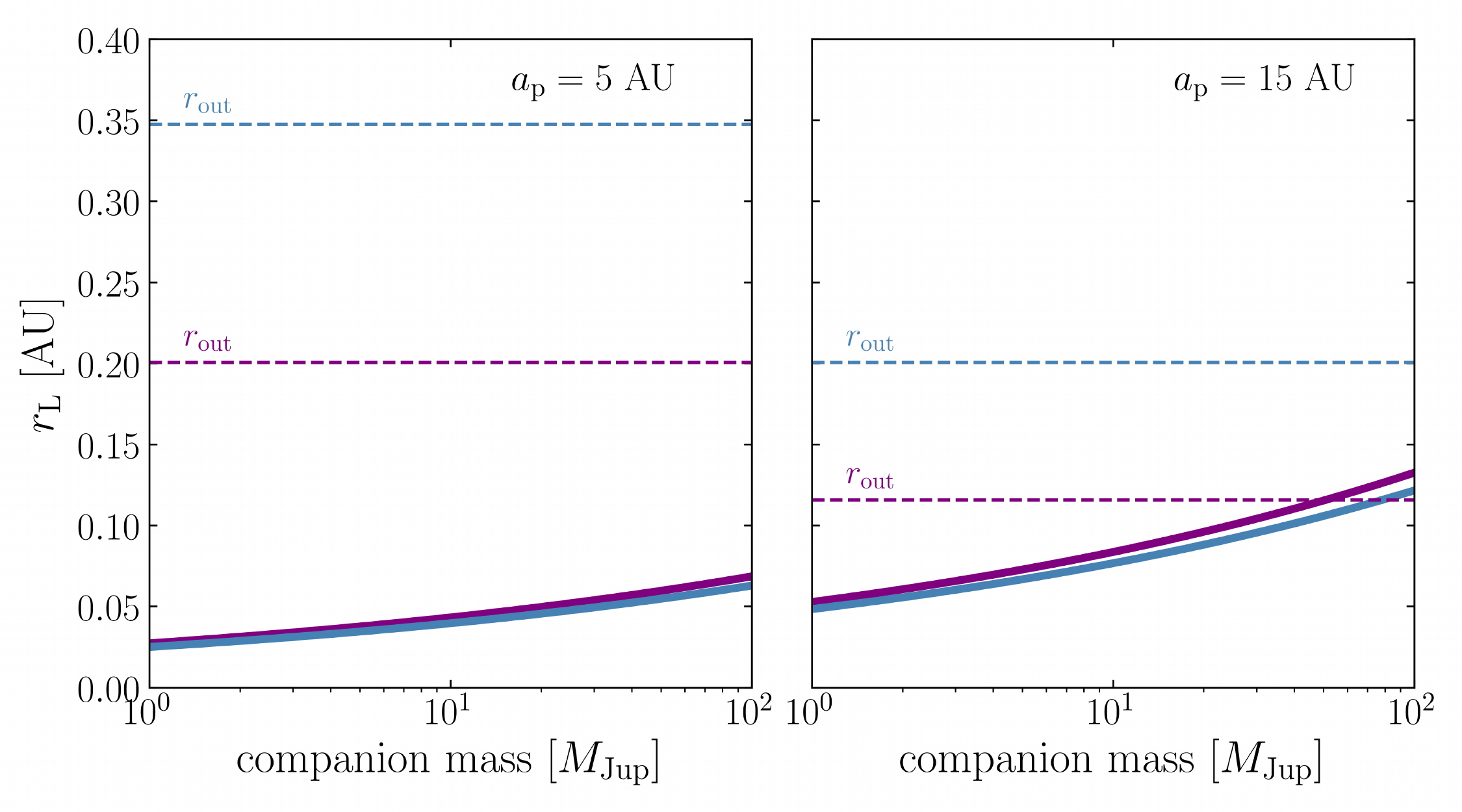}
    \caption{
    The Laplace radius (eq.~\ref{eq:rL}; solid lines) as a function of companion mass $\Mp$, compared with the observationally constrained radius of J1407b's disk $\rout$ (eq.~\ref{eq:rout}; dashed lines), for circular ($\ep = 0$; blue lines) and eccentric ($\ep = 0.5$; purple lines) companion orbits.  For a transit signal to be produced by J1407b's disk, we require $\rL \sim \rout$ (see also Fig.~\ref{fig:surface}).  Here, $\Ms = 0.9 \, \Msun$, $\Rp = 1 \, \Rj$, $t_{\rm ecl} = 54 \, {\rm days}$, $f_{\rm ecl} = 0$, and $J_2 = 0.5$.
    }
    \label{fig:laprad-mass}
\end{figure*}

A number of challenges remain regarding the stability of the putative disk orbiting J1407b.  In order to achieve the high transverse velocities implied by the rapid changes in the J1407 light curve, either the companion J1407b must be on a tight orbit ($\ap<6$ au), or we are observing the periastron passage of the companion on a very eccentric orbit ($\ep>0.7$; \citealt{kenworthy2014mass, mentel2018constraining}). Whether J1407b lies on an eccentric orbit which transits at periastron, or on a tight circular orbit, the J1407b disk would fill a significant fraction (70-100\%) of the companion's Hill sphere \citep{mentel2018constraining}. A disk orbiting retrograde was found by \citet{rieder2016constraints} to survive longer ($\sim$110,000 years) around a subtellar-mass (60-100 $\Mj$) companion on an 11-year orbit than a disk orbiting prograde, but this poses the additional challenge of explaining the formation of a retrograde circumplanetary disk. 

Furthermore, while the flat, inclined and extended best-fitting ring model of \citet{kenworthy2015modeling} reproduces many of the features in the observed J1407 light curve, it does not reproduce them all. The authors suggest that better fits to the data may be achieved by a model that allows for additional degrees of freedom for the rings, such as warping or precession. A disk around J1407b that warps to the Laplace surface, such as we investigate in this work, may produce better fits to the data. It may also be a solution to the challenges in stability faced by its flat counterpart. 

The remaining question is then the maximum size of the circumsecondary disk around J1407b.  We review the order-of-magnitude estimates of \cite{mamajek2012planetary} constraining the size of J1407b's disk, but note similar estimates can be used to constrain properties of other tentative circumsecondary disks. The time over which J1407 is eclipsed is $t_{\rm ecl} \sim 54 \, {\rm days}$ \citep{mamajek2012planetary}. 
For this eclipse time to be due to a circumsecondary disk, we must have
\begin{equation}
    t_{\rm ecl} \sim \frac{2 \rout}{v_{\rm orb}} = \frac{2 \rout}{n_{\rm p} \ap F(\ep, f_{\rm ecl})},
\end{equation}
where $v_{\rm orb}$ and $f_{\rm ecl}$ are the orbital velocity and true anomaly of the companion during eclipse, $\ep$, $n_{\rm p} = \sqrt{G \Ms / \ap^3}$, and $\ap$ are the companion's eccentricity, mean-motion and semi-major axis, while $\rout$ is the outer radius of the disk. Here,
\begin{equation}
    F(\ep, f_{\rm ecl}) = \sqrt{\frac{1 + 2 \ep \cos f_{\rm ecl} + \ep^2}{1 - \ep^2}}
\end{equation}
is how much $v_{\rm orb}$ can be increased/decreased from a non-zero $\ep$. Observations constrain the companion's orbital period $P_{\rm orb}$ and eclipse duration $t_{\rm ecl}$, so we solve for $\rout$:
\begin{align}
    &\rout \sim F(\ep, f_{\rm ecl}) \, \Big( \frac{\pi G \Ms}{4 P_{\rm orb}}\Big)^{1/3} \, t_{\rm ecl} 
    \nonumber \\
    &= 0.080 \, F(\ep, f_{\rm ecl}) \Big(\frac{\Ms}{0.9 \, \Msun} \Big)^{1/3} \Big( \frac{30 \, {\rm yrs}}{P_{\rm orb}}\Big)^{1/3} \Big( \frac{t_{\rm ecl}}{30 \, {\rm days}}\Big) \ {\rm au}  .
    \label{eq:rout}
\end{align}

For a warped disk to be a feasible model, the tilted disk region's size must be large enough to produce the $\sim$54-day transit seen by \cite{mamajek2012planetary}. Since the inclined disk's size $r_{\rm inc}$ is $r_{\rm inc} \sim \rL$ (see Fig.~\ref{fig:surface}, see also \citealt{SchlichtingChang(2011)}), we require $\rout \sim \rL$ for a disk warped along the secondary's Laplace surface to explain the observations.  Figure \ref{fig:laprad-mass} calculates the Laplace radius $\rL$ (eq.~\ref{eq:rL}) of a warped disk surrounding a companion to J1407 as a function of the companion's mass $\Mp$ for two $\ap$ values, and compares $\rL$ to the observationally constrained outer disk radius $\rout$. 
We find that a warped circum\textit{planetary} disk ($M_{\rm p} \lesssim 13 \, \Mj$) around J1407b cannot extend to large enough radii to explain the observations. An eccentric companion orbit creates a larger discrepancy between $\rL$ and $\rout$. A substellar companion on a large, circular orbit could produce a long enough transit signal, though it would be inconsistent with the transverse velocities required to produce the rapid changes observed in J1407's light curve. For a warped disk to explain J1407b's transit, additional forces must work to increase the radius of $r_{\rm inc}$ to be greater than $\rL$, which we discuss in the next subsection.




\subsection{Theoretical Uncertainties and Future Work} 

This work studies the dynamics of collisionless circumplanetary debris disks, neglecting internal processes that may modify the orbits of massless particles from which the disk is composed.  Radiation pressure is widely known to have a significant impact on circumstellar debris disks \citep[e.g.][]{LeeChiang(2016)}, and can modify the Laplace surface for circumplanetary disks \citep{Tamayo(2013)}.  Additional forces within the disk -- such as disk self-gravity \citep{Ward(1981),zanazzi2016extended}, along with bending waves \citep[i.e.][]{zanazzi2017inclination} and viscous torques \citep{TremaineDavis(2014)} in hydrodynamical disks -- can modify the equilibrium warp profile.  These additional external or internal forces/torques can potentially increase the radius of the disk's tilted region $r_{\rm inc}$ to be greater than $\rL$ (eq.~\ref{eq:rL}).  In particular, \cite{zanazzi2016extended} found disk self-gravity can increase $r_{\rm inc}$ to become comparable to the observationally constrained $\rout$ for J1407b's disk (see Sec.~\ref{sec:Obs_rout} for discussion), but the disk mass required to do so ($M_{\rm disk} \gtrsim {\rm few} \ 10^{-3} \, \Mp$) may cause the disk to become gravitationally unstable.  Moreover, since this disk is optically thick \citep{kenworthy2015modeling}, particle collisions are likely relevant \citep[e.g.][]{WisdonTremaine(1988),ReinPapaloizou(2010),Latter(2012)}.  We leave investigations of these effects to future works.

We argue ivection resonance \citep{xu2019exciting} causes the inclination excitation observed in the inner region of the circumplanetary disk in our integrations (see Sec.~\ref{subsec:ivec} for discussion), but further theoretical work is needed to understand its role in detail.  We believe it was not investigated before because previous studies on satellites around massive planets had the satellite semi-major axis $r$ slowly increase due to tidal dissipation \citep[e.g.][]{ToumaWisdom(1998),CukBurns(2004)}.  In our preliminary integrations reducing the disk's inner truncation radius $r_{\rm in}$ (which we do not show), we find the location of evection resonance ($r\lesssim 0.1 \, \rL$) to be interior to ivection resonance ($r\approx 0.2 - 0.5 \, \rL$).  Ivection resonance may be particularly important for exomoons, since the resonant semi-major axis $r_{\rm res}$ (eq.~\ref{eq:r_res}) will increase as the planet migrates inward through its natal protoplanetary disk \citep{Spalding(2016)}.

To date, all studies examining the transit signatures of circumplanetary disk systems assume a flat disk, or coplanar rings, misaligned with the planet's orbital plane.  For the tentative detections of extended circumplanetary disks \citep[e.g.][]{mamajek2012planetary,Osborn(2017),Saito(2019),Rappaport(2019)}, our study shows such disks are likely to be significantly warped.  Studies of transit signatures from warped circumplanetary disks would be of interest for these extended systems.


\section{Conclusions}
\label{sec:conclusions}

Motivated by the tentative detection of extended circumplaneary/circumsecondary disk/ring systems, we explore the dynamical stability of particles orbiting near the equilibrium surface (Laplace surface) set by the balance between the tidal torque from the host star with the torque from the spinning planet's mass quadrapole, using $N$-body integrations.  By integrating test particle orbits drawn from distributions close to the Laplace surface, we simulate the evolution of a collisionless circumplanetary debris disk system.  We perform our simulations for three planetary obliquities ($\bgp=20^{\circ}$, $50^{\circ}$, $80^{\circ}$) and two values of eccentricity ($\ep=0$, $0.5$). Our main findings are as follows:

\begin{enumerate}
    \item For all planet obliquities ($\bgp=20^{\circ}$, $50^{\circ}$, $80^{\circ}$) and eccentricities ($\ep=0,0.5$), much of the warped circumplanetary disk remains stable over the course of our integrations ($\sim 3-16 \, {\rm Myr}$; Sec.~\ref{sec:results}).
    \item Two dynamical resonances/instabilities can potentially carve gaps in the circumplanetary disk: the Lidov-Kozai mechanism (Sec.~\ref{subsec:LK}) and ivection resonance (Sec.~\ref{subsec:ivec}).  Such gaps could be detectable if the circumplanetary disk system transits its host star, and would not necessarily imply the existence of exomoons (Sec.~\ref{sec:exomoons}).
\end{enumerate}

We also investigate if a disk orbiting the hypothesized J1407 companion could be misaligned by J1407b's mass quadrupole, and extend to large enough radii to match the observed eclipse duration seen in \cite{mamajek2012planetary}. We find the disk may become misaligned at sufficiently large distances from J1407b only for a substellar-mass ($\Mp \gtrsim 50\, \Mj$) companion on a large ($a\gtrsim15$ au) nearly-circular orbit.  Additional forces are needed to extend the misaligned region of the disk around J1407b to explain J1407's eclipse duration.

Shortly after this work was published, \cite{Kenworthy(2020)} used the Atacama Large Millimeter/submillimeter Array (ALMA) to detect potential sub-millimeter continuum emission around the star J1407, and attempted to detect thermal emission from the J1407b exoring system.  \cite{Kenworthy(2020)} detected optically thin dust emission orbiting a substellar object, consistent with a ring system if J1407b lies on an unbound orbit.  If this detected substellar object is indeed the J1407b ring system, our stability constraints must be modified to account for an impulsive encounter between the disk and host star.  Our results remain applicable to the general stability of extended circumsecondary disks.

Photometric techniques may soon bring direct observations of circumplanetary disks to reality, providing opportunity to put to the test the regions of instability we predict in this work. Discussions of circumplanetary (or circumsecondary) debris disks, exorings, or even exomoons should consider the geometry of the Laplace surface.


\section*{Acknowledgements}

We thank Cristobal Petrovich for useful discussions.  JJZ thanks Dong Lai and Matija Cuk for useful discussions on evection resonance.  JS was funded by a Canadian Institute for Theoretical Astrophysics (CITA) Natural Sciences and Engineering Research Council (NSERC) Undergraduate Student Research Award (USRA), while JJZ is a CITA postdoctoral fellow.  

We would like to acknowledge this sacred land on which the University of Toronto operates. It has been a site of human activity for 15,000 years. This land is the territory of the Huron-Wendat and Petun First Nations, the Seneca, and most recently, the Mississaugas of the Credit River. The territory was the subject of the Dish with One Spoon Wampum Belt Covenant, an agreement between the Iroquois Confederacy and Confederacy of the Ojibwe and allied nations to peaceably share and care for the resources around the Great Lakes. Today, the meeting place of Toronto is still the home to many Indigenous people from across Turtle Island and we are grateful to have the opportunity to work in the community, on this territory.

\textit{Software:} 
\texttt{astropy} \citep{robitaille2013astropy, price2018astropy},
\texttt{jupyter notebook} \citep{Kluyver:2016aa}, 
\texttt{matplotlib} \citep{hunter2007matplotlib}, 
\texttt{numPy} \citep{walt2011numpy},
\texttt{pandas} \citep{mckinney-proc-scipy-2010},
\texttt{rebound} \citep{rein2012rebound},
\texttt{sciPy} \citep{2020SciPy-NMeth}.

\textit{Data Availability:} The data underlying this article will be shared on reasonable request to the corresponding author.


\bibliographystyle{mnras}
\bibliography{SpeedieZanazziv3} 



\appendix

\section{Total integration time break-down}
\label{sec:app:integration}

As mentioned in Section~\ref{sec:simulations}, our integration times for test particles within each disk segment (semi-major axis bin) are set by computational constraints.  Table~\ref{tab:appendix} lists the break-down of integration times by particle segment, for the $\ep = 0$, $\bgp = 50^\circ$ integration; the other integrations displayed in Figures~\ref{fig:betapanel-ep0} and~\ref{fig:betapanel-ep05} are similar.

\begin{table*}
\caption{Total integration time by disk segment for the $\ep=0$, $\bgp=50^{\circ}$ integration. Columns are: test particle index (\#), the segment of the disk it belongs to (seg.), its semi-major axis ($a$, in units of $\rL$), the total integration time of that segment (${\rm T}_{\rm tot}$ $10^6$ [yr]), and how many circumplanetary test particle orbits to which ${\rm T}_{\rm tot}$ equates to ($10^6$ [orbits]). All particles in Segment 1 are shown; only the first is shown for segments 2-49; first and last are shown for Segment 50. Segments that are \textit{not} integrated to 16 Myr are indicated by *.}
\begin{center}
\begin{tabular}{|*{11}{c|}}  
\hline
\multicolumn{6}{|c}{Segments 1, 2-16} & \multicolumn{5}{c|}{Segments 17-50} \\ \hline

\# & seg. & $a$ [$\rL$] & ${\rm T}_{\rm tot}$ $10^6$ [yr] & $10^6$ [orbits]  & \ & \# & seg. & $a$ [$\rL$] & ${\rm T}_{\rm tot}$ $10^6$ [yr] & $10^6$ [orbits] \\ \hline
0   & 1 *  & 0.2001 & 3.2646  & 85.669 &  & 320 & 17 & 0.7761 & 16.0000 & 54.971 \\
1   &   & 0.2016 & 3.2646  & 84.737 &  & 340 & 18 & 0.8121 & 16.0000 & 51.356 \\
2   &   & 0.2038 & 3.2646  & 83.370 &  & 360 & 19 & 0.8480 & 16.0000 & 48.123 \\
3   &   & 0.2055 & 3.2646  & 82.327 &  & 380 & 20 & 0.8840 & 16.0000 & 45.215 \\
4   &   & 0.2071 & 3.2646  & 81.368 &  & 400 & 21 & 0.9199 & 16.0000 & 42.593 \\
5   &   & 0.2090 & 3.2646  & 80.239 &  & 420 & 22 & 0.9560 & 16.0000 & 40.204 \\
6   &   & 0.2107 & 3.2646  & 79.274 &  & 440 & 23 & 0.9920 & 16.0000 & 38.039 \\
7   &   & 0.2125 & 3.2646  & 78.278 &  & 460 & 24 & 1.0280 & 16.0000 & 36.056 \\
8   &   & 0.2142 & 3.2646  & 77.373 &  & 480 & 25 & 1.0640 & 16.0000 & 34.243 \\
9   &   & 0.2163 & 3.2646  & 76.199 &  & 500 & 26 & 1.0999 & 16.0000 & 32.578 \\
10  &   & 0.2180 & 3.2646  & 75.356 &  & 520 & 27 & 1.1360 & 16.0000 & 31.040 \\
11  &   & 0.2197 & 3.2646  & 74.461 &  & 540 & 28 & 1.1719 & 16.0000 & 29.622 \\
12  &   & 0.2218 & 3.2646  & 73.421 &  & 560 & 29 & 1.2080 & 16.0000 & 28.305 \\
13  &   & 0.2234 & 3.2646  & 72.630 &  & 580 & 30 & 1.2441 & 16.0000 & 27.083 \\
14  &   & 0.2254 & 3.2646  & 71.670 &  & 600 & 31 & 1.2803 & 16.0000 & 25.942 \\
15  &   & 0.2268 & 3.2646  & 70.988 &  & 620 & 32 & 1.3161 & 16.0000 & 24.892 \\
16  &   & 0.2287 & 3.2646  & 70.092 &  & 640 & 33 & 1.3524 & 16.0000 & 23.896 \\
17  &   & 0.2305 & 3.2646  & 69.271 &  & 660 & 34 & 1.3876 & 16.0000 & 22.992 \\
18  &   & 0.2324 & 3.2646  & 68.438 &  & 680 & 35 & 1.4237 & 16.0000 & 22.123 \\
19  &   & 0.2342 & 3.2645  & 67.651 &  & 700 & 36 & 1.4596 & 16.0000 & 21.312 \\
20  & 2 *  & 0.2362 & 4.1823  & 85.582 &  & 720 & 37 & 1.4966 & 16.0000 & 20.527 \\
40  & 3 *  & 0.2721 & 5.1233  & 84.788 &  & 740 & 38 & 1.5318 & 16.0000 & 19.823 \\
60  & 4 *  & 0.3080 & 6.0529  & 83.177 &  & 760 & 39 & 1.5677 & 16.0000 & 19.146 \\
80  & 5 *  & 0.3442 & 7.0285  & 81.764 &  & 780 & 40 & 1.6045 & 16.0000 & 18.491 \\
100 & 6 *  & 0.3801 & 8.2411  & 82.587 &  & 800 & 41 & 1.6392 & 16.0000 & 17.907 \\
120 & 7 *  & 0.4159 & 9.5217  & 83.383 &  & 820 & 42 & 1.6759 & 16.0000 & 17.323 \\
140 & 8 *  & 0.4520 & 10.7267 & 82.919 &  & 840 & 43 & 1.7123 & 16.0000 & 16.773 \\
160 & 9 *  & 0.4881 & 12.0153 & 82.766 &  & 860 & 44 & 1.7474 & 16.0000 & 16.270 \\
180 & 10 * & 0.5240 & 13.4099 & 83.041 &  & 880 & 45 & 1.7841 & 16.0000 & 15.771 \\
200 & 11 * & 0.5600 & 14.8241 & 83.097 &  & 900 & 46 & 1.8190 & 16.0000 & 15.319 \\
220 & 12 & 0.5959 & 16.0000 & 81.695 &  & 920 & 47 & 1.8559 & 16.0000 & 14.864 \\
240 & 13 & 0.6320 & 16.0000 & 74.791 &  & 940 & 48 & 1.8921 & 16.0000 & 14.440 \\
260 & 14 & 0.6680 & 16.0000 & 68.827 &  & 960 & 49 & 1.9273 & 16.0000 & 14.046 \\
280 & 15 & 0.7041 & 16.0000 & 63.614 &  & 980 & 50 & 1.9655 & 16.0000 & 13.639 \\
300 & 16 & 0.7400 & 16.0000 & 59.039 &  & 999 & 50 & 1.9986 & 16.0000 & 13.301 \\ \hline
\end{tabular}
\end{center}
\label{tab:appendix}
\end{table*}


\section{Further Investigations of Particle Stability}

\subsection{Longitude of Ascending Node}
\label{sec:app:Omega}

In addition to variations in particle inclination $\beta$ and eccentricity $e$, we explore variations in the longitude of ascending node $\Omega$. A particle can become misaligned with the Laplace surface if $\Om$ moves out of the Laplace plane (or, synonymously, if its angular momentum vector moves out of the plane spanned by the planet's orbit angular momentum and spin vectors).

\begin{figure}
	\includegraphics[width=\columnwidth]{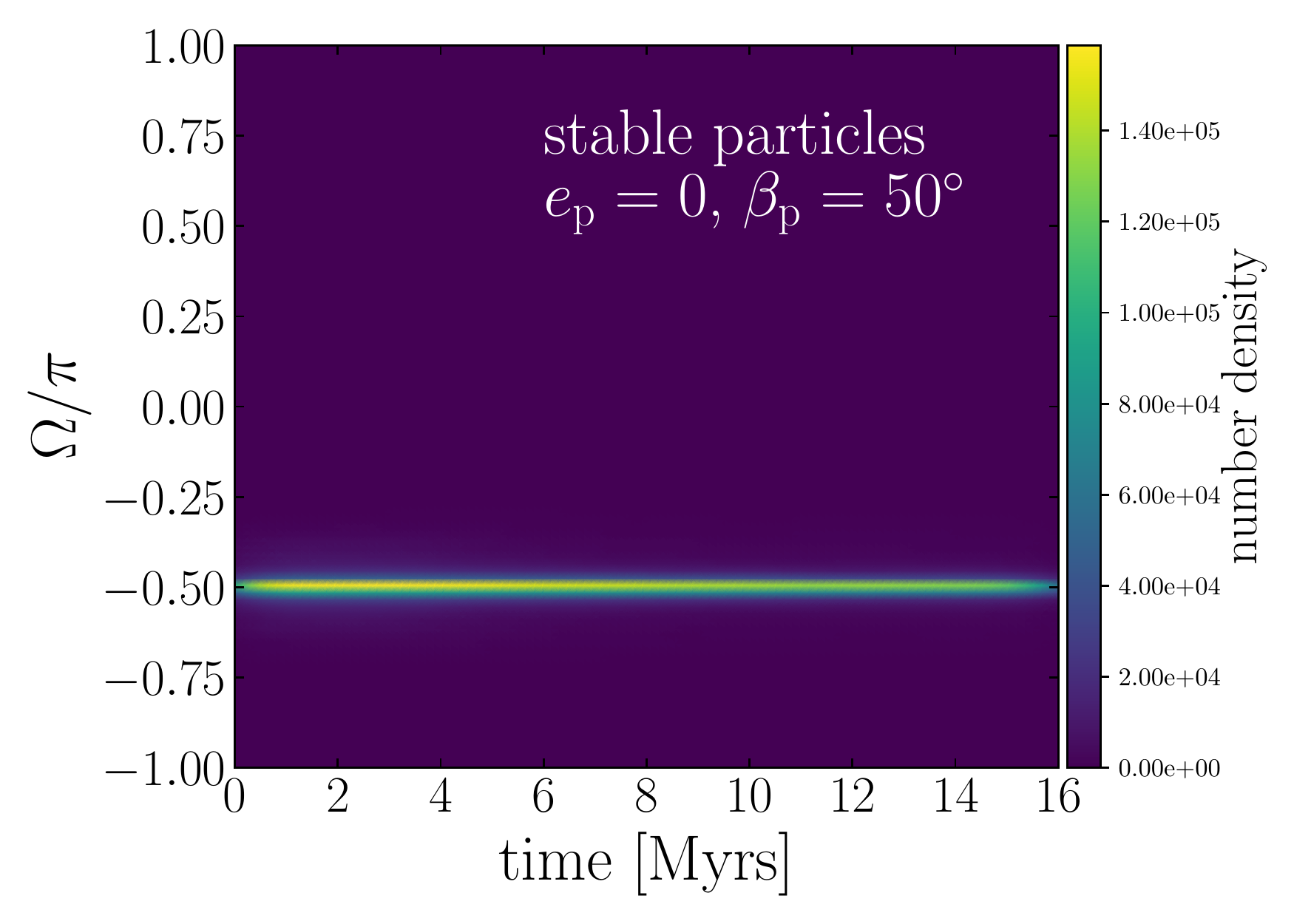}
	\includegraphics[width=\columnwidth]{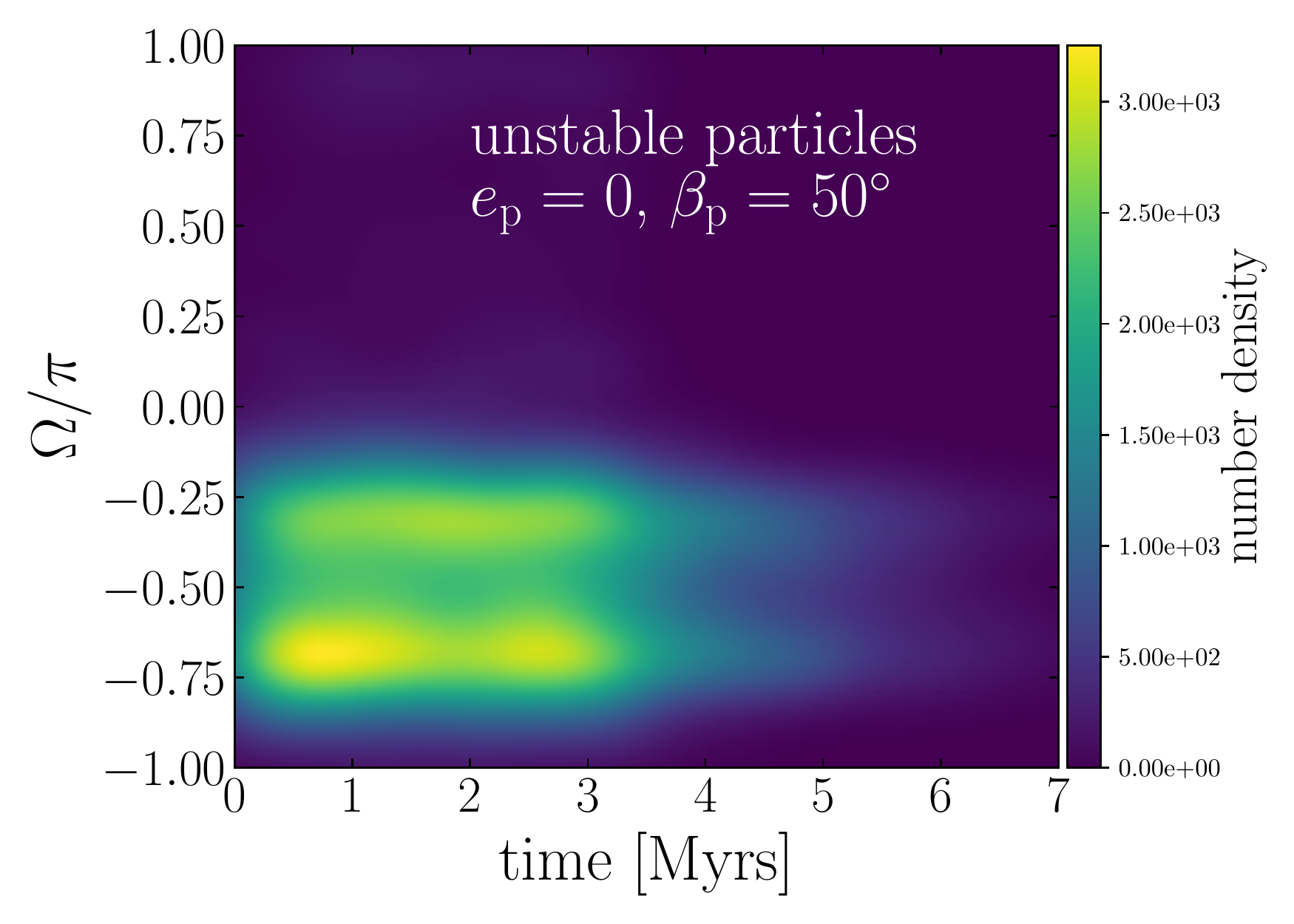}
    \caption{ Evolution of test particle longitude of ascending node $\Omega$ of stable (top panel) and unstable (bottom panel) particles, from the $\bgp= 50 ^{\circ}$, $\ep=0$ integration (middle row of Fig. \ref{fig:betapanel-ep0}).  The vast majority of stable test particle $\Om$ values lie close to the Laplace surface's longitude of ascending node $\Om = -\pi/2$.
    }
    \label{fig:omega-v-time}
\end{figure}

In Figure \ref{fig:omega-v-time} we show the $\Omega$ evolution of all stable and unstable particles in the $\ep=0$, $\bgp =50^{\circ}$ integration. The vast majority of particles lie near the Laplace surface's longitude of ascending node ($\Om = -\pi/2$), while the unstable particles do exhibit nontrivial deviations from $\Om = -\pi/2$. The number density in the bottom panel of Fig. \ref{fig:omega-v-time} falls off with time, and the x-axis only extends to 7 Myr instead of the fully 16 Myr, because we performed the integrations for a fixed number of orbits. This corresponds to less physical time for the innermost particles, which are also the (ivection) unstable particles in this case. We have verified that the other five integrations have similar results.

\subsection{Ivection-Unstable Regions}
\label{sec:app:ivection}

\begin{figure*}
	\includegraphics[width=18cm]{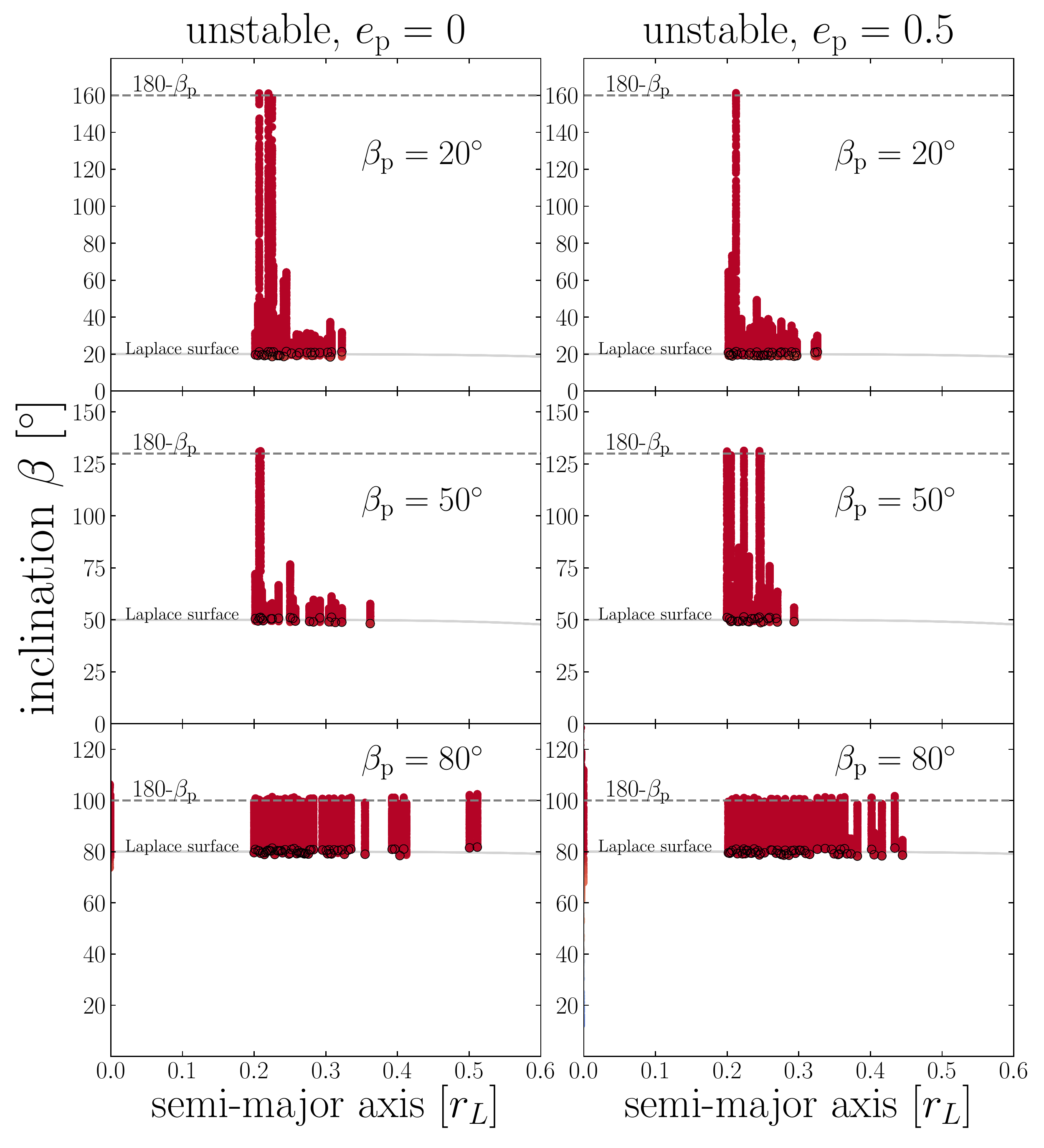}
    \caption{The inclinations $\bg$ of unstable particles in Fig.~\ref{fig:betapanel-ep0} (left panels) and Fig.~\ref{fig:betapanel-ep05} (right panels), focusing on the inner regions of the disk susceptible to ivection instability ($r < 0.6 \ \rL$).  The full range of $\bg$ values are displayed, alongside the Laplace surface inclination $\bgp$ (solid grey line) and $180^\circ - \bgp$ (dashed grey line).  All particles have negligible eccentricity over the course of the integration ($e[t] \lesssim 0.01$).
    }
    \label{fig:ivection-zoom}
\end{figure*}

Figure~\ref{fig:ivection-zoom} examines the full inclination dependence $\bg(r)$ of the disk regions susceptible to ivection resonance, examining the inner regions ($r < 0.6 \, \rL$) of the unstable regions in Figures~\ref{fig:betapanel-ep0} and~\ref{fig:betapanel-ep05}.  Particles with smaller semi-major axis $r$ tend to have larger $\bg$ excitations than large $r$ values (save for the $\bgp = 80^\circ$ integration), with particles near $r \approx 0.2 \, \rL$ excited to retrograde inclinations ($\bg \approx 180^\circ - \bgp$).
We speculate the $r$ dependence of the $\bg$ excitation is related to the test particle's proximity to the ivection (and possibly eviction, \citealt{ToumaWisdom(1998)}) resonance location(s).  It is not clear to the authors why the $\bgp = 80^\circ$ integration has $\bg$ excitations which don't seem to depend on $r$.

\begin{figure*}
	\includegraphics[width=18cm]{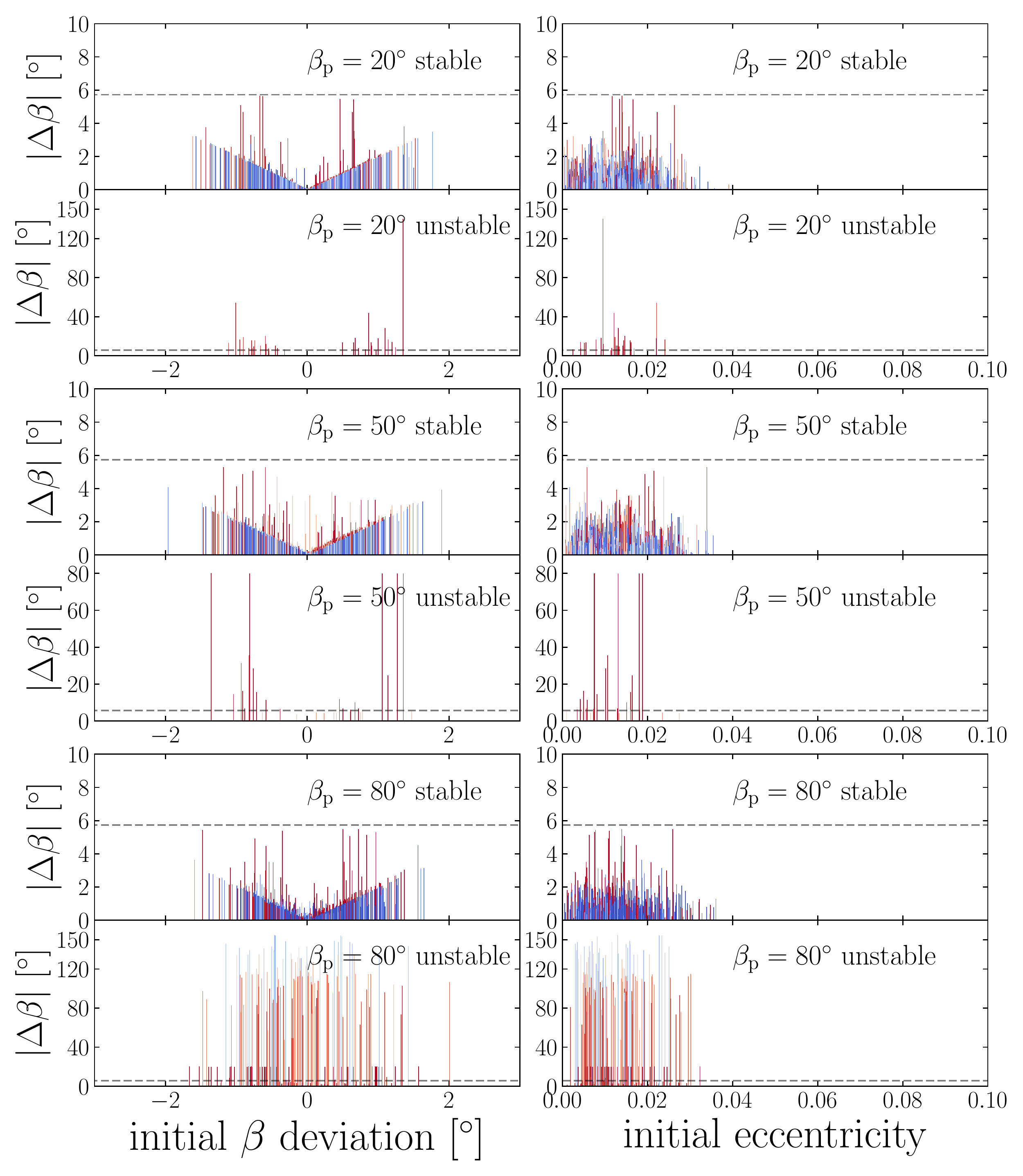} 
    \caption{Maximum particle inclination excitation $|\Delta \bb|$ as a function of initial inclination deviation and initial eccentricity for the three $\ep=0.5$ integrations (see Fig. \ref{fig:betapanel-ep05}). An initial $\bb$ deviation of $0^{\circ}$ and $e$ of $0$ corresponds to a particle initialized directly on the Laplace surface. In all panels, the horizontal dashed grey line corresponds to the stability condition $|\Delta \bb| = |\bb[t]-\bb[0]| < 0.1$ rad $= 5.73^{\circ}$.
    }
    \label{fig:beta-vs-perturbation}
\end{figure*}

Figure \ref{fig:beta-vs-perturbation} presents a preliminary exploration into \textit{why} there is a variation in inclination excitation amongst the ivection-unstable particles. For the three $\ep=0.5$ integrations (Fig. \ref{fig:betapanel-ep05}), we plot test particle inclination as a function of initial $\bb$ deviation from the Laplace surface, and as a function of initial eccentricity. The results for the three $\ep=0$ integrations look virtually identical.

The stable and unstable distribution of particles in the righthand column of Figure \ref{fig:beta-vs-perturbation} are very similar, suggesting little or no dependence of inclination excitation on eccentricity. In other words, a higher initial eccentricity does not improve a particle's chance of becoming unstable and undergoing ivection resonance.

The unstable particles in the lefthand column of Figure \ref{fig:beta-vs-perturbation}, however, tend to be distributed $\gtrsim 0.5^{\circ}$ away from 0 (above or below the Laplace surface). The stable particles in this same column, particularly those at large semi-major axes (shown in blue) display a linear dependence of $|\Delta \bb|$ on the initial $\bb$ deviation due to nodal precession from the host star's tidal torque.

\bsp	
\label{lastpage}
\end{document}